\documentclass[12pt]{article}
\usepackage{amsmath}
\usepackage{amssymb}
\usepackage{epsfig}
\usepackage{cite}
\usepackage{color,colordvi}

\begin{document}
\vspace{0.01cm}
\begin{center}
{\Large\bf  
Quantum  Compositeness of Gravity:  Black Holes, AdS and Inflation
} 

\end{center}

\vspace{0.1cm}

\begin{center}

{\bf Gia Dvali}$^{a,b,c}$ and {\bf Cesar Gomez}$^{a,e}$\footnote{cesar.gomez@uam.es}

\vspace{.6truecm}


{\em $^a$Arnold Sommerfeld Center for Theoretical Physics\\
Department f\"ur Physik, Ludwig-Maximilians-Universit\"at M\"unchen\\
Theresienstr.~37, 80333 M\"unchen, Germany}


{\em $^b$Max-Planck-Institut f\"ur Physik\\
F\"ohringer Ring 6, 80805 M\"unchen, Germany}

%


{\em $^c$Center for Cosmology and Particle Physics\\
Department of Physics, New York University\\
4 Washington Place, New York, NY 10003, USA}

{\em $^e$
Instituto de F\'{\i}sica Te\'orica UAM-CSIC, C-XVI \\
Universidad Aut\'onoma de Madrid,
Cantoblanco, 28049 Madrid, Spain}\\

\end{center}


\begin{abstract}
\noindent  
 
{\small

 Gravitational backgrounds, such as black holes,  AdS, de Sitter  and inflationary universes, should be viewed as 
composite of $N$
soft constituent gravitons.  It then follows that such systems are close to quantum criticality of graviton Bose-gas to Bose-liquid 
transition. 
Generic properties of the ordinary metric description, including geodesic motion or 
particle-creation in the background metric, emerge as the large-$N$ limit of quantum scattering of constituent longitudinal gravitons. 
We show that this picture correctly accounts 
for physics of large and small black holes in AdS, as well as reproduces well-known inflationary 
predictions for cosmological parameters. However, it anticipates new effects not captured by the standard semi-classical treatment. In particular, we predict observable corrections that 
are sensitive to the inflationary history way beyond last 60 e-foldings. 
 We derive an absolute upper bound on the number of e-foldings,
 beyond which neither de Sitter nor inflationary Universe can be approximated by a semi-classical 
 metric.  However, they could in principle persist in a new type of {\it quantum eternity}  state.   
 We discuss implications of this phenomenon for the cosmological constant problem.  }

\end{abstract}

\thispagestyle{empty}
\clearpage

\section{Introduction}

In  \cite{Nportrait, Quantum, Hair} we have developed a quantum compositeness  theory of 
black holes.   The different aspects of this proposal and some related ideas were discussed in 
\cite{P1, string-hair, P2, P3, P4, P5, P6, scramble, P7, P8, P9}.
 In this description a black hole of a classical size $R$ is a composite {\it quantum} system of many soft gravitons of  similar wavelength.  We have discovered that this composite system 
exhibits properties of a  Bose-Einstein condensate ``frozen" at a quantum critical point.  The critical point corresponds to a quantum phase transition between what we can call graviton-Bose-gas and graviton-Bose-liquid phases. 
 Because of this proximity to quantum criticality, even a seemingly-classical macroscopic system 
in reality is intrinsically-quantum. The effects due to quantum compositeness become extremely 
important and must be taken into account.  These effects are not captured by any of the standard 
analysis, since these approaches are blind to the compositeness of the would-be-classical background.  
 
  In \cite{Nportrait}  (see also \cite{P1}) we have also suggested that compositeness treatment should be generalized to other gravitational systems, such as the maximally-symmetric cosmological space-times.  In particular, we showed that when describing AdS space in  a similar fashion, the occupation number of gravitons 
 coincides with what would be the central charge of a CFT, computed according to  the AdS/CFT prescription \cite{Malda}.  This remarkable coincidence  indicates that compositeness and quantum criticality 
 of gravitational   backgrounds may be the underlying reason for an 
effective holographic description. 

In the present paper we shall investigate the compositeness picture of gravity from several  different perspectives.    
  
   $~~~$
   
    In what follows we adopt the key concept of \cite{Nportrait}: 
   
   $~~~$
   
{\it Gravitational systems, such as black holes, AdS, de Sitter  or other cosmological spaces  represent composite entities 
 of microscopic quantum constituent gravitons of  wave-length set by the characteristic classical size $R$
 (i.e., the curvature radius) of the system. }
 
 $~~~$
 
   Obviously, for a black hole $R$ has to be understood as the gravitational radius $R_{BH}$, whereas for 
   AdS/ dS or cosmological spaces, as the corresponding curvature (
   Hubble ) radius,  $R_{AdS}, R_{dS},  R_{H}$ respectively.

  In the present paper we shall systematically apply our treatment to  AdS, de Sitter and inflationary spaces. First, we conduct several consistency checks, reproducing known phenomena as a particular 
 approximation of our approach.  However, we go beyond the simple consistency check and uncover new effects that are not captured by standard treatments. These new effects are due to the quantum compositeness of the gravitational background. 
 
    We first discuss how the known aspects of the ordinary curved-metric description of gravity emerge
    as the $N=\infty$ limit of our picture. We show, how the motion of a probe source in an effective 
   classical background metric is recovered from its quantum  scattering at the constituent 
   off-shell gravitons.    The entities that  play the crucial role in recovery of  the classical metric, are matrix elements 
   of the form,
   \begin{equation}
    \langle N + 1| a^+ |N\rangle  \, \sim \,
 {\sqrt{\hbar}  \over R}  \sqrt{N +1} \,,~~~   \langle N| a^+ a |N\rangle  \, \sim \,
 {\hbar  \over R^2}  N \,~~~\, ...\, , 
 \label{matrixIntroduction}
  \end{equation} 
   where $|N\rangle$ is the quantum state of the gravitational background of 
   occupation number $N$ of gravitons of wavelength $R$ and $a^+$ and $a$ are creation and annihilation 
   operators of the constituent off-shell gravitons.   These matrix elements are non-vanishing in the semi-classical 
   limit and recover the motion in the background classical metric. 
   
    As we shall see,  transitions in which all initial and final constituent gravitons are off-shell only contribute 
    into the recovery of the classical metric, without particle creation. Whereas, processes in which some of the 
    final constituent gravitons become on-shell, amount to particle creation.   

 In particular, we explain in this way the well-known effects of Hawking \cite{hawking} and 
 Gibbons-Hawking\cite{deSitter}  radiation, as well as the scalar (Mukhanov-Chibisov \cite{slava1}) and tensor 
 (Starobinsky \cite{Inflation2}) modes of inflationary density perturbations.  Moreover, we explain certain aspects of this phenomenon, that in semi-classical treatment appears to be rather mysterious. 
 Namely, why only space-times with globally un-defined time exhibit the phenomenon of 
 particle-creation.  In our description only such space-times correspond to critical quantum systems in which the 
 constituent gravitons carry non-zero frequencies,  and thus, are able to produce on-shell particles in their re-scattering.  In static space-times with globally-defined time the Bose-gas of constituent gravitons is either far form criticality and/or  the constituent gravitons have zero frequencies and cannot lead to creation of on-shell particles in their interactions.  
 
   The important lesson we are distilling is that the particle-creation is {\it not} a vacuum process, and consequently, 
   there is no maximal entanglement created in a single emissions act.  This goes in contrast with the 
   standard interpretation of Hawking radiation according to which a particle-pair is created out of the vacuum, and thus, is maximally 
   entangled.  In reality the particles are not created out of vacuum, but rather they result from re-scattering of existing 
  constituent gravitons. 
  Only in $N=\infty$ limit 
   such a particle-creation process can be naively interpreted as a vacuum pair-creation process. This is because in this limit the 
   background effectively becomes classical. It gains infinite capacity of emitting particles and, therefore, becomes eternal.  
  To such an eternal background  the entanglement can be attributed arbitrarily, since it can 
   never be measured during any finite time.  However, for any finite-$N$ background, 
  would be a severe mistake to  assume a creation of a maximal entanglement in a single particle 
  emission act.  It takes a number of steps of order $N$ to create a maximal entanglement among the constituents.   
  
   Next, we shall apply our picture to different cosmological spaces.  Among other things, we shall describe physics of both small and large black holes in AdS in simple physical terms. 
    
    Implementation of compositeness ideas to the case of de Sitter and inflationary universes, allows us to reproduce the well known inflationary predictions, such as the scalar \cite{slava1} and tensor \cite{Inflation2}
 perturbations, as particular (large-$N$) approximations of the full quantum picture.
  Most importantly, we discover some new effects which substantially change our view about the 
 consistency, as well as observability, of the very early cosmological history of our Universe.  
 
  The key novelty for inflation is that compositeness acts as a {\it quantum} clock that 
  imprints measurable effects into cosmological observables.  These effects are cumulative and gather the information throughout the entire duration of inflation.  
   Hence, in our picture, a given Hubble patch carries quantum information about the entire history 
   of inflation,  way beyond the last $60$ e-foldings. This quantum information  is not redshifted away 
  by the expansion and can be read-off after the end of inflation within the same Hubble patch.  
   Obviously, the effects we are talking about are unaccessible in semi-classical treatment and have been missed in all the previous analysis. 
   
   The physical origin of these effects is nevertheless very transparent.  In our treatment, de Sitter or an inflationary 
   Hubble patch of a classical radius $R_H$, is quantum mechanically interpreted as a reservoir of a finite number of gravitons and inflatons, i.e., as a Bose-Einstein condensate.  The occupation number of gravitons is, 
     \begin{equation}
     N\, =  \, (R_{H}  / L_P)^{2} \, , 
  \label{NN}
 \end{equation}
 where $L_P$ is the Planck length, which is related to Newton's constant as $L_P^2 \, = \, \hbar \,  G_N$. 
   Just as in the black hole case, this condensate is close to quantum criticality.  As a result,  it undergoes 
 an intense quantum depletion, which gradually decreases the occupation number of gravitons and 
 inflatons in the course of inflation and works towards emptying the reservoir.  
 
   The novelty, in comparison with the black hole case, is that the graviton depletion is assisted by the 
 inflaton background, which itself represents a Bose-gas of occupation number $N_{\phi} \, \gg \, N$.
 The key point is that the inflaton-to-graviton ratio is controlled by the classical inflationary slow-roll 
 parameter,\footnote{See  \cite{Review}  for 
a review on inflation and standard definitions.}  $\epsilon \equiv (V' /VL_P)^2 \hbar $, as   
\begin{equation}
   {N \over N_{\phi}} \, = \, \sqrt{\epsilon}  \, , 
  \label{epsilon}
\end{equation}
where $V$ is the inflaton potential and  $V' \, \equiv \, dV/d\phi$.  
  Thus,  slower the inflaton rolls  higher is  the ratio $N_{\phi}/N$  and the condensed gravitons have 
 increasing number of inflaton partners to re-scatter-at and jump out of the condensate.   
  As a result the depletion is enhanced by a factor  $N_{\phi} \over N$ as compared to the black hole case. 
   As we shall see, this factor also explains the enhancement of the scalar mode in density perturbations
   \cite{slava1,slava2} relative to the tensor one \cite{Inflation2}. 
Moreover, this depletion of the reservoir produces a measurable effect sensitive to the entire duration of inflation. 

   The remarkable thing is that, since the quantum depletion rate is controlled by the classical slow-roll  
  parameter,  there is a consistency upper bound on how slow inflation can be and therefore the number of e-foldings 
 is bounded from above \footnote{Some interesting previous attempts, within the standard semi-classical treatment,  to set a bound on the number of e-foldings based on assigning finite entropy to de Sitter can be found in \cite{Bounds1,Bounds2, Bounds3}.  These approaches do not resolve quantum compositeness of the gravitational background and therefore cannot account for the effects we are discussing. Correspondingly, our bound 
 has a different physical meaning and, as we shall see, turns out to be more severe.}.  
 
  In particular, the depletion rate becomes infinite for the exact  de Sitter limit $\epsilon =0$.  This excludes de Sitter  as a consistent limit of a slow-roll inflation and also puts an upper bound on the possible number of e-foldings in any theory of inflation that admits an approximate semi-classical description throughout the history.   
 
  This is a very subtle point and we would like to avoid a mis-interpretation.   This 
  statement only applies to de Sitter obtained as a limit of slow-roll inflation and it  
    does not touch upon the de Sitter without extra light scalar degrees of freedom, such as in case of e.g.,  
  a pure cosmological term, or a local minimum of the potential, as in  the  ``Old" inflationary scenario \cite{Inflation1}. However, such a ``pure"  de Sitter is also severely constrained by our composite
  picture, and cannot exist eternally as a state which admits an approximate  description in terms of 
  a classical metric.  We cannot exclude the possibility that it can continue existence eternally in a quantum 
  state that admits no approximate classical description.  This interesting option will be discussed in the text.

    The physics that we have just summarized can be foreseen from the following master equation

    \begin{equation}
 {\dot{N} \over N}   \, = \,  H \, \left ( \epsilon  \, - \, {1 \over \sqrt{\epsilon}} {1 \over N} \right ) \, , 
  \label{masterintro}
  \end{equation}
where $H \, \equiv \, R_H^{-1}$ is the Hubble parameter.

 The first term  in the r.h.s. of this equation describes the purely-classical time-evolution of the graviton occupation 
 number due to the classical increase of the Hubble radius with cosmic time.  This term of course survives 
 in the classical limit $\hbar \, = \, 0$. 
 
  The second term describes the decrease of $N$ due to quantum depletion of the background gravitons.   
 This is the term that puts a consistency upper bound on the duration of inflation and excludes the 
 de Sitter limit $\epsilon=0$.  The fact that $\epsilon$ cannot be arbitrarily small is obvious. 
 To understand why,  we can first take it to be small enough that the first term becomes irrelevant.   In this regime $N$ 
  evolves exclusively due to quantum depletion, 
 \begin{equation} 
    \dot{N} |_{quantum} \, = \, - H/\sqrt{\epsilon} \, , 
    \label{onlyquantum}
 \end{equation}
 or expressing everything in terms of occupation numbers,  
 \begin{equation} 
    \dot{N} |_{quantum} \, = \, - {1 \over \sqrt{N}L_P} { N_{\phi}\over N} \, . 
    \label{onlyquantumN}
 \end{equation}
This is very similar to our black hole  depletion equation \cite{Nportrait} (see below), apart from an extra enhancement factor 
$N_{\phi} \over N$.  
  In some sense, whenever the dynamics is dominated by quantum depletion,  the Hubble patch enters into a black hole-type regime, except with exceedingly high depletion rate.    
  Consequently, the standard computation of e-foldings in terms of $\epsilon$ is no longer possible. 
  Instead, the duration of inflation is determined by the time that takes to deplete order one fraction 
  of $N$.  If this time is shorter than the classically evaluated number of e-foldings, the theory is inconsistent. 
   To illustrate the point let us integrate the above equation for one Hubble time, during which we treat 
  $\epsilon$ and $H$ as constants. 
    For the change we get $\Delta N \equiv N_{in} \, - \, N_{fin} \, =\, 1/\sqrt{\epsilon}$. 
     Thus, the maximal number of Hubble times, consistent quantum-mechanically, is 
     ${\mathcal N}_e^{quantum} \, = \, N \sqrt{\epsilon}$. Indeed, this is the time needed to deplete the
     entire reservoir! 
           This number must be consistent with the classically evaluated 
   number of e-foldings ${\mathcal N}_e^{class}$,  which gives a unique consistency 
   bound. 
     \begin{equation}
    \epsilon \, > \, N^{-2/3} \, .
    \label{boundepsilon} 
  \end{equation}

    The precise expression in terms of ${\mathcal N}_e$ is determined by the dependence of  ${\mathcal N}_e^{class}$ on $\epsilon$. 
   For example,  for $V\, = \, m^2\phi^2$ inflation \cite{Chaotic},  ${\mathcal N}_e^{class}  \, = \, 1/\epsilon$, and we get the following consistency bound,
 
    \begin{equation}
    {\mathcal N}_e \,  < \, N^{{2\over 3}} \, .
   \label{boundintro}
  \end{equation} 

 In order to understand the robustness of this bound, imagine that as a would-be counter attempt we choose $\epsilon \, = \, 1/N^2$.  A naive classical person would 
 conclude that the number of e-foldings is ${\mathcal N}_e^{class} \, = \, N^2$, but in reality 
 quantum depletion empties the condensate during ${\mathcal N}_e^{quantum} \, = \, 1$. This mismatch means that such a theory is simply inconsistent, viewed as a description of an approximately semi-classical  system.   
  The absolute bound is given by (\ref{boundintro}). 

  This bound excludes any potential that slopes to a positive cosmological constant and puts a severe restriction on 
  how slow the decrease of potential energy could be.

 The bound translated as a constraint on the classical potential energy, has the following form, 
   \begin{equation}
    \left({M_PV' \over V} \right )^2 \, < \, {1 \over \hbar} \left ({M_P^4\over V} \right )^{2/3}\,.   
    \label{generalbound}
    \end{equation}
From here it is clear that smaller is the 
potential energy slower it is permitted to evolve.  In particular, observational constrains on  time-evolution of dark energy satisfy this bound. 

 It is important to stress that the bound is not an artifact due to the breakdown of some perturbation theory
 and cannot be removed by any re-summation.  The reason is that it is coming from the enhanced 
 phase space of the depletion,  due to the excessive occupation number of inflatons in the 
 Bose-condensate which catalyze graviton depletion.  This factor cannot be removed by re-summation 
 since the number of channels is physical. In this respect, the situation with the de Sitter limit is similar 
 to the depletion of black holes in the presence of large number of massless particle species\cite{species}. 
  
   When the number of massless species is infinite,  theories  become inconsistent with the existence of  black holes, since the number of particle-emission channels becomes infinite  and this fact cannot be saved by perturbative re-summation.  
  Similarly,  in the composite picture the de Sitter limit is inconsistent, because it supplies an infinite number of assistant  "depletors" and 
  the rate of particle-production blows up.

  We shall show that  the depletion term in (\ref{masterintro}) can contribute to a potentially-measurable 
  cumulative effect in the cosmological observables, such as amplitude of density perturbations 
  and tensor-to-scalar ratios.   Hence, by precision measurement of these parameters one can 
  scan the history of our Hubble patch beyond the last $60$ e-folds.  
   
    Another cumulative quantum effect that is also sensitive to the entire history to the Hubble patch is 
   the  {\it quantum entanglement}  of the constituent gravitons.   It was argued \cite{scramble} that a near-critical condensate of gravitons generate entanglement very efficiently, with minimal time-scale of the order $\sim R$ln$(N)$.  If this is also the case for the Hubble graviton reservoir \footnote{See \cite{Susskind1},\cite{F} for an attempt to extend the notion of scrambling \cite{Scrambling1, Scrambling2} to cosmology.} then after this time-scale it becomes an entangled 
    quantum system. This entanglement is carried by the depleted gravitons and in this way is imprinted 
    in density perturbations.  
       For  realistic values of the inflationary parameters, the entanglement  time-scale is within
    $30$ e-foldings. This gives an exciting possibility to measure the age  of inflation by detecting 
    entanglement among the depleted gravitons.  How to fish out this information from the measurements is an interesting question to be studied in the future. 
    
     We would like to stress,  that none of the correction we are talking about have to do with 
   either  trans-Planckian gravity or other UV-sensitive regions.  All our effects are taking place due to interactions of very soft gravitons 
     for which gravity is extremely weak and quantum gravity effects are computable and are under an excellent control.

   Finally, we would like to suggest that our picture can provide a quantum foundation of 
   what is sometimes referred  to as "holography". 

Our point is \cite{Quantum} that near quantum criticality a system of $N$ soft gravitons, no matter how macroscopic, exhibits an enormous degeneracy ($\sim exp(N)$) of states corresponding to nearly-gapless collective Bogoliubov modes, which behave as 
nearly conformal with an effective central charge determined by the inverse gravitational coupling of the original
constituent gravitons. It is a simple fact that inverse gravitational coupling $\alpha^{-1}$ in any space-time dimension scales 
as area in Plank units.  Hence the reason for our adoption of the term  "holography".

     This also naturally explains why a holographic description (in our sense) is only available for certain  gravitational systems and not for others.  These are the systems for which the quantum effects are very important, but they have mistakenly been treated classically. 
    These are the systems that are close to the quantum critical point and for which the underlying 
    quantum compositeness is very important.   For example, black holes, AdS and 
    cosmological spaces.

  Holographic description is subjected to corrections due to compositeness  of the gravitational system. 
  These corrections, although suppressed by fractions of $N$, should not be confused with
  corrections to 't Hoofts planar limit on the gauge theory side \cite{tHooft} (if for a particular system  such a description is available). 
    Corrections we are talking about are fundamentally different.  In particular, they have a cumulative 
  effect and become extremely important at later times as it is the case for cosmology or black hole entanglement. 

The logic flow of the paper is the following. After briefly reviewing the compositeness approach to black holes we sketch the formalism underlying the quantum description of geometry as the large $N$ limit of certain graviton condensates. The key ingredient is the replacement of a classical background curved metric with  
the quantum notion of a condensate composed out of off-shell (longitudinal) gravitons. 
 Different phenomena of the conventional metric description are then recovered as large-$N$ limits of various quantum scattering 
processes with the participation of these constituent gravitons.  For example, we identify processes that 
reproduce the geodesic motion in a background classical metric, as well as the processes of depletion that are 
responsible for the particle creation.  

  As we shall see, the on-shell dispersion relation for resulting metric fluctuations is determined ({\it collectively }) by the condensate state itself. The quantum depletion depends on the particular type of constituent gravitons defining the condensate and sets the time properties of the emergent metric as well as the dynamics of  quantum particle production. 
  We work out the examples of de Sitter and anti de Sitter space-times and explain the difference in 
  particle-creation.   
  
  Next,  we extend the condensate formalism to the inflationary cosmology where we reproduce the standard results on quantum fluctuations in terms of quantum depletion. We work out compositeness corrections and determine the bound on maximal number of e-foldings.
   These results motivate us to visualize the cosmological constant problem from a very different perspective. 
   
   Finally we consider the case of AdS and black holes with AdS boundary conditions in the graviton condensate formalism. 
   
    Throughout the paper, irrelevant factors of order one will be ignored, but signs and important relative factors 
    will be followed carefully.    
    
\section{Black Holes}

   Before proceeding to AdS and Inflationary spaces,  we wish to briefly review 
 some ingredients of our black hole quantum portrait \cite{Nportrait} that will be useful later for drawing analogies as 
 well as differences with other spaces.   According to our picture a black hole of classical radius 
 $R_{BH}$ must be viewed as a collection of constituent gravitons of wavelength $\lambda \, = \, R_{BH}$. 
 This automatically fixes its occupation number according to (\ref{NN}), 
  \begin{equation}
  N \, = \, {R_{BH}^2 \over L_P^2} \,.
   \label{NBH}
  \end{equation}
  The notion of the above occupation number of soft gravitons as black hole constituents  was introduced in \cite{class1} and in 
  quantum portrait it represents a measure of classicality \footnote{In a different context, the same measure of classicality is also adopted in 
  \cite{Ramy}.}.    
   
 Thus, the quantum gravitational coupling among the gravitons, $\alpha \, \equiv \, L_P^2 /R^2$,  satisfies,
   \begin{equation}
   \alpha =  {1 \over N}\,. 
   \label{criticalpoint}
  \end{equation} 
 From this relation it follows that the system is described as a self-sustained bound-state, which 
 exhibits all the properties of a Bose-gas at the critical point of a quantum phase transition. 
 This critical point corresponds to $\alpha N =1$ and separates, what one could call, the Bose-gas and Bose-liquid phases. 
 
  To understand the  significance of this point, note that  for  $\alpha N\, < \, 1$ the gravitational attraction 
  would not be strong enough to keep the gravitons together.  Contrary, for $\alpha N\, > \, 1$   the gravitational attraction among the constituents would induce instability towards quantum collapse. 
  So the critical point is unstable. But, precisely because of criticality, there is an enormous degeneracy
  of collective (Bogoliubov) excitations that are nearly gapless.  As a result, 
  there are two important quantum effects that determine the evolution of the system. 
  
  One is the quantum depletion, which is very strong at the critical point.  The reason for the depletion is that 
  constituent gravitons re-scatter and are pushed out of the condensate. At the same time the condensate 
  shrinks because of the quantum collapse. As a result the gravitons continuously leak out of the condensate at the rate given by the following time evolution equation
  \begin{equation}
  \dot{N} \, = \, -  {1 \over \sqrt{N} L_P}  \, + \, {\mathcal O}(N^{-1}) \, .
  \label{leakage}
  \end{equation}
  Thus, due to quantum depletion the graviton condensate looses one graviton of wave-length $\lambda\, = \, \sqrt{N} \, L_P \, = \, R_{BH}$,  per emission time $t\, = \, \sqrt{N} \, L_P$.   Notice, that unlike the standard 
  Hawking's semi-classical computation, the emission of particles in our case is not a vacuum process, in which 
 particles are created in a fixed background geometry.  
  Rather, in our case the particle emission from a black hole is due to depletion of gravitons that are already 
  pre-existing in the condensate. As a result, the condensate back-reacts and recoils.  
   The two descriptions match in the semi-classical limit,  
  \begin{equation}
  L_P \, = \, 0, ~~~\hbar \, = \, {\rm fixed}, ~~~  R_{BH} \, = \, {\rm fixed} \, .  
  \label{semiclasslimit}
  \end{equation}
 In this limit,  our quantum depletion process  indeed 
  reproduces the Hawking's thermal evaporation of temperature  
  $T_H \, = \, \hbar /R_{BH}$. 
   Obviously, in this limit $N \, =\, \infty$ and the quantum compositeness of the black hole is not resolved. 
   However, as we have shown, for finite $N$ there are extremely important corrections. These  corrections per each emission are suppressed by $1/N$ and naively look unimportant for large $N$.  
   But, this is a wrong intuition, since the number of emissions is  also growing as $N$. So 
   the cumulative effect of the corrections is very important and leads to the total breakdown of semi-classical 
   approximation over time-scales comparable to the black hole half life-time $t \, \sim \, N^{3/2} L_P$.  
   
    After this time scale the black hole can no longer be regarded -- even approximately -- as a classical system. 
   As we shall see, there is a full analog of this phenomenon for the inflationary Universe, which also 
   becomes fully quantum after finite time.  
         
    Notice, that  if the theory contains extra $N_{species}$ light particle species of mass $m \, \ll \, \hbar/R_{BH}$, the depletion rate is increased by a factor $N_{species}$,     
     \begin{equation}
  \dot{N} \, = \, -  {1 \over \sqrt{N} L_P} \, N_{species}  \, ,
  \label{leakagemany}
  \end{equation}
   due to the existence of extra depletion channels.  
  Hence, the quantumness of the black hole increases with increasing number of species. 
  As we shall see, this effect too finds a counterpart in the inflationary case. In particular, 
 the  presence of  inflaton quanta increases the depletion rate, which explains the excess of scalar 
 curvature perturbations over the tensor mode. 
    
     Thus, as we shall see, both equations (\ref{leakage})  and (\ref{leakagemany}) have analogs  
  in the inflationary Universe.

    Finally, we would like to make some general comments with the goal of putting some typical black hole paradoxes in the perspective of compositeness. It is probably not surprising that 
we can make a black hole in pure gravity simply putting together a certain amount $N$ of soft gravitons. If we are thinking about large black holes,  the description in terms of constituents looks at first sight very easy to handle.  Indeed the constituents can be chosen to be arbitrarily weakly-coupled by taking their wave-length
large.  From this perspective it is hard to expect anything dramatic as long as their number is below the 
black hole formation threshold $N \, \ll \, \alpha^{-1}$. 

   The first black hole puzzle from this compositeness point of view appears when we try to understand what actually happens when you add the "last" soft constituent and you suddenly pass from a simple aggregation of weakly coupled constituents into a black hole. From the classical General Relativity  point of view this is a very special moment where the space-time arena where the constituents were originally living changes dramatically. Thus, the obvious question is, what happens at this point to the many-body quantum system we have used to build up the black hole. The answer to this basic question is remarkably simple: the many body system of gravitons becomes critical and undergoes a quantum phase transition! The key reason is also simple to understand. The constituents are interacting, but the collective  interaction strength,  $\alpha N$,
that appears as the coupling of the corresponding Hamiltonian also changes and reach the critical point
$\alpha N = 1$
 at the precise moment of black hole formation. 

From the classical General Relativity point of view what appears as extraordinary after black hole formation are the properties of the resulting space time, and more specifically, the nature of space-time inside the black hole.  Thus the next obvious question is, what is the counterpart of the geometry in the compositeness picture. The answer is  that the 
connection between the two descriptions is through the scattering amplitude. Namely, the quantum scattering of a probe particle at the condensate is equivalent to a geodesic motion in the background 
classical metric. However, it is extremely important to understand that this description becomes less and less 
accurate as the time goes on, and the black hole constituents become entangled. 
The important fact is, that at the quantum critical point an intense generation of quantum entanglement starts taking place.  After finite time, the constituents that we were naively putting together to create the black hole become  quantum-mechanically maximally entangled. 

 The effect of entanglement  is intrinsically finite-$N$ effect and correspondingly the time that takes 
 the black hole constituents to become maximally entangled is larger for large  $N$. Due to this, the entanglement effect  is impossible 
to capture in the standard semi-classical description since in this case $N \, = \, \infty$ and generation of entanglement takes 
infinite time.   
 We shall discuss this effect in more details below when we shall make the 
   study of a similar effect for the inflationary Universe.  
      
      Before we start to apply our picture to cosmological spaces we will present the basis of the compositeness formalism. To do that we will work out explicitly the graviton condensate representation of maximally symmetric space times. Incidentally this detailed construction might  provide the clue -- from the graviton condensate point of view--  to unveil the black hole inner geometry .
  
  \section{Emergence of the Curved Geometry} 
  
  We now wish to show how the concept of a curved classical metric emerges in our picture as a result 
  of quantum scattering. 
  
   Since we substitute the {\it classical} notion of a curved geometry by the {\it quantum} notion 
 of large graviton occupation number on a Minkowski vacuum,  the curved metric must 
 only appears as an effective emergent description.  Indeed,  we shall demonstrate that the quantum scattering 
 of a probe particle at the graviton condensate,  in the limit 
 (\ref{semiclasslimit}) of large occupation number, reproduces the  geodesic motion in the background classical metric.  
 
    For definiteness, we shall consider a graviton condensate that in the classical limit  reproduces de Sitter space.  For simplicity, we shall limit ourselves to the analysis that in the classical limit 
  reproduces the leading process and the first sub-leading (non-linear) correction to the motion in a 
  small-curvature (weak field) background.  Under {\it small curvature}  we mean a situation when the curvature radius is much larger than the characteristic wave-length of the probe. 
 We shall show, that  this semi-classical picture fits the quantum scattering of a probe of  wave-length shorter than the characteristic 
  wave-length of the graviton condensate.   As we shall see, this type of scattering reproduces a geodesic motion in a  
 small-curvature classical metric background.

   In order to fix the language, let us first define the classical problem to which we map our large-$N$ limit. 
 Classically, a metric  created by an energy-momentum source $T_{\mu\nu}$ can be obtained by 
 iteratively solving   Einstein's equation in weak field expansion of the metric, 
 \begin{equation}
   g_{\mu\nu} \, = \,  \eta_{\mu\nu} \, + \, h_{\mu\nu} \, + \, h_{\mu\nu}' \, +\, ... \,.
   \label{weakexpansion}    
 \end{equation}
   Here $\eta_{\mu\nu}$ is the flat Minkowski metric, whereas $h_{\mu\nu}$  and $h_{\mu\nu}'$ are obtained in
higher order iterations.  Namely, $h_{\mu\nu}$ represents a solution of  the linearized Einstein' equation
\begin{equation}
  {\mathcal E} h_{\mu\nu} \, = \, G_{N} T_{\mu\nu}   \, , 
 \label{lineareinstein}
 \end{equation} 
 where ${\mathcal E} h_{\mu\nu} $ is the linearized Einstein's tensor, 
 \begin{equation}
 {\mathcal E} h_{\mu\nu} \, \equiv \, \square h_{\mu\nu} \, - \,\eta_{\mu\nu} h \, -\,  \partial_{\mu} \partial^{\alpha} h_{\alpha\nu} \, - \,  \partial_{\nu} \partial^{\alpha} h_{\alpha\mu} \, +  \, 
  \eta_{\mu\nu}  \partial^{\alpha} \partial^{\beta} h_{\alpha\beta} \, + \, 
 \partial_{\mu} \partial_{\nu} h \,,  
\label{einsteintensor}
\end{equation}
   whereas, $h_{\mu\nu}'$ represents the first correction due to cubic self-interaction of gravity, and 
   satisfies the equation, 
 \begin{equation}
  {\mathcal E} h_{\mu\nu}' \, = \, G_{N} T_{\mu\nu}(h)   \, ,  
 \label{secondorder}
 \end{equation} 
  where, 
   \begin{equation}
  T_{\mu\nu}(h)   \, \equiv  \,  - \,  h^{\alpha\beta}  \partial_{\nu} \partial_{\mu} h_{\alpha\beta} \, + \, ... \,,
 \label{tensorh}
 \end{equation} 
  is the effective energy momentum tensor of gravity evaluated on the first order perturbation 
  $h_{\mu\nu}$.  For economy, we shall not display the full expression.  One can continue this iterative process to arbitrary high orders.  Resuming the entire series we would recover the metric 
  $g_{\mu\nu}$ that satisfies the fully non-linear Einstein's equation. This type of expansion, with a suitable choice of gauge, can be done anywhere in a small curvature region.  
  
   In order to make contact with our quantum picture we should first promote $h_{\mu\nu}$ into 
   a quantum field, 
\begin{equation}
 h_{\mu\nu}  \, \rightarrow   \, {L_P \over \sqrt{\hbar}} \, \hat{h}_{\mu\nu}  \, ,  
 \label{graviton}
 \end{equation} 
 where ${L_P \over \sqrt{\hbar}}$ stands for ensuring the canonical dimensionality.   
  Then $h_{\mu\nu}$ encodes the classical limit of processes that do not involve graviton self-interactions, 
  whereas $h_{\mu\nu}'$ carries information about the cubic self-scattering,  and so on. 
  
   Before going into a  full quantum picture, let us note that within the classical (or semi-classical) description 
   there are two equivalent ways of thinking about the motion of a probe of energy-momentum 
   $\tau_{\mu\nu}$ in a classical gravitational field produced by the source $T_{\mu\nu}$. 
   
   One interpretation, is to  first compute the classical  gravitational field $g_{\mu\nu}$ produced by 
   $T_{\mu\nu}$, either exactly or order by order in weak field expansion (\ref{weakexpansion}), 
   and then evaluate a geodesic motion of the source $\tau_{\mu\nu}$ from its coupling 
    to the derived classical metric, 
    \begin{equation}
   g_{\mu\nu}  \,\tau^{\mu\nu} \,.   
 \label{metriccoupling}
 \end{equation} 
 
     The second interpretation is to think about the motion of $\tau_{\mu\nu}$ as the result of scattering 
   between $\tau_{\mu\nu}$ and $T_{\mu\nu}$ via virtual graviton exchanges. 
   
    For example, in first order approximation we can think of motion of $\tau_{\mu\nu}$ 
  in the linear classical metric  
   \begin{equation}
  \int d^4x \,  h_{\mu\nu}  \,\tau^{\mu\nu} \,,    
 \label{classmotion}
 \end{equation} 
 where  $h_{\mu\nu}$ is the solution of the linear Einstein equation (\ref{lineareinstein}). Alternatively, we can  think of the same process as the result of one-graviton exchange amplitude between $\tau_{\mu\nu}$ and $T_{\mu\nu}$, 
   \begin{equation}
 s_1 \, = \,  {L_P^2 \over \hbar} \, \int d^4x d^4\tilde{x} \, \tau^{\mu\nu} (x) \, \Delta_{\mu\nu, \alpha\beta} (x \, - \, \tilde{x}) \,T^{\alpha\beta}(\tilde{x}) \, ,    
 \label{linearexchange}
 \end{equation} 
 where  $\Delta_{\mu\nu, \alpha\beta} (x \, - \, \tilde{x}) \, \equiv \, \langle \hat{h}_{\mu\nu}(x) , \hat{h}(\tilde{x})_{\alpha\beta}  \rangle$  a graviton propagator.   For example, in de Donder  gauge it can be written as, 
   \begin{equation}
 \Delta_{\mu\nu, \alpha\beta} \, = \,{ {1 \over 2}( \eta_{\mu\alpha}  \eta_{\nu\beta} \,  + \, 
 \eta_{\mu\beta}  \eta_{\nu\alpha})\, - \, {1\over 2}   \eta_{\mu\nu} \eta_{\alpha\beta}
 \over \square}  \, .  
 \label{propagator}
 \end{equation} 
  Of course, there is nothing surprising in the fact that the two seemingly different languages, one classical and another quantum, give the same result.  In fact, both languages are classical, 
  since the quantity $G_N \, \equiv \, {L_P^2 \over \hbar}$ is nonzero in the limit 
  $\hbar \, = \, 0$.   Of course, this equivalence holds to all orders in weak field expansion. 
  For example, second order correction to the motion in the classical metric is given by 
    \begin{equation}
 s_2 \, = \, \int d^4x d^4\tilde{x} \, \tau^{\mu\nu} (x) \, \Delta_{\mu\nu, \alpha\beta} (x \, - \, \tilde{x}) \,T^{\alpha\beta}(h(\tilde{x})) \, ,    
 \label{nonlinearexchange}
 \end{equation}  
  where $T^{\mu\nu}(h(\tilde{x})$ is the energy momentum tensor of the linear perturbation, $h_{\mu\nu}$, given 
  by (\ref{tensorh}).  Obviously, this 
  correction is second order in $G_N$, since  $T^{\mu\nu}(h(\tilde{x})$ is bilinear in 
  $h_{\mu\nu}$ which is first order in $G_N$. 
  
     The possibility of tree-level Feynman diagram representation of the classical metric, such as 
   Schwarzschild,  is well known \cite{Duff}.  Our first task will be to give an analogous expansion for 
   cosmological spaces, and then to show how this 
   classical expansion can be understood as the large-$N$ limit of quantum processes that take 
   into account the composition of the background.   
  
   Although in the above formalism the gravitational field is classical, the probe  $\tau_{\mu\nu}$ can be 
   either classical or quantum. In the latter case we are working in semi-classical approximation,  
 and the transitions between the initial and the final states of the probe can be found perturbatively in the form of $s_n$ matrix elements, 
 \begin{equation}
 \langle f | s_{n} |i\rangle \, . 
 \label{transit}
 \end{equation}
For example, $|i\rangle$ and $|f\rangle$ states can describe initial and final states of an incident photon  (of energy momentum tensor  $\tau_{\mu\nu}$)  scattering in the gravitational field of the sun 
(of energy momentum tensor  $T^{\mu\nu}$).

    The bottom line is, that any quantum picture that in the semiclassical limit reproduces the above weak field expansion has a correct classical Einsteinian limit. 
   We wish to demonstrate this for our composite picture, according to which a would-be classical   
  geometry has to be viewed as a quantum state $|N_k\rangle $ with large occupation numbers  $N_k$ of gravitons 
  with wave-numbers $k$, and with the peak of the distribution being at the characteristic curvature radius of the system.   The motion in the classical background geometry is then reproduced in large-$N$ limit 
  as scattering between the probe particle and the constituent gravitons.  
  
   For definiteness, we shall establish this dictionary for reproducing the de Sitter geometry, which classically would amount to the motion into a background metric $g_{\mu\nu}$ satisfying the Einstein equation with the source $T_{\mu\nu} \, = \, g_{\mu\nu} H^2$, where $H$ is the Hubble constant. 
  We shall  consider evolution of short wave-length probes, with the wavelength 
  $\lambda$ shorter than the Hubble radius  $H^{-1}$.  
  For such processes, we can reliably use the weak field expansion and treat the process perturbatively, by  
first  considering the motion on a background linear metric, 
$h_{\mu\nu}$, satisfying the equation 
 \begin{equation}
  {\mathcal E} h_{\mu\nu} \, = \, \eta_{\mu\nu}  \, H^2   \, ,  
 \label{cc}
 \end{equation} 
 and treating effects of non-linearities as higher order corrections.  

It is easy to check \cite{giajustinstefan}, that the above equation admits several gauge-equivalent solutions that 
 describe de Sitter space for short time-scales and distances. For example, 
 \begin{equation}
 h_{00} \, = \, h_{0i} \, = \, 0 \, ~~ \, h_{ij} \, = \, H^2 (t^2\delta_{ij}\, + \, n_in_j r^2) \, ,    
 \label{linearDS} 
 \end{equation}  
 where $n_i \, \equiv \, x_i/r \,,~ \, r\equiv \, \sqrt{x_j^2}$, is a linearlized 
de Sitter metric in closed FRW slicing, and we have set an over-all numerical coefficient to one.  
  This represents an approximation of a full non-linear metric, 
  \begin{equation}
 ds^2 \, = \, - \, dt^2 \,  + \, {\rm cosh}(Ht) \left ( {dr^2 \over 1\, - \, H^2r^2} \, + \, r^2 d\Omega^2 \right )\, ,    
 \label{nonlinearDS} 
 \end{equation}  
 for  $Hr \, \ll \, 1$ and $Ht \, \ll \, 1$.

   We now need to re-interpret this geometry as the large-$N$ limit of a quantum state with 
   large occupation number of constituent gravitons.  We shall refer to these constituents 
   as longitudinal gravitons.   We chose the gauge in order to reproduce  the de Sitter geometry 
   in closed FRW slicing (\ref{linearDS}).  Then, in large $N$ limit we need only to consider 
   $h_{ij}$ components being non-zero.  
   
   Let us expand the classical metric perturbation $h_{ij}$ into the Fourier harmonics,   
        \begin{equation}
   h_{ij} \, = \, \int d^4k \, e^{ikx} \, ( I_{ij} b_{I,k} \, + \, L_{ij} b_{L,k} )   \, + \, 
  e^{-ikx} \, ( I_{ij} b^*_{I,k} \, + \, L_{ij} b^*_{L,k} ) \, . 
   \label{operatorexp}
   \end{equation}  
 Here we have introduced two orthogonal tensor projectors, $I_{ij} \, \equiv  \, \delta_{ij} \, - \, n_{i}n_{j}$ and $ L_{ij} \, \equiv \, n_{i}n_{j}$, and $b,b^*$ are  expansion coefficients.   The field $h_{ij}$  does not satisfy any free wave equation. 
 Correspondingly,  the wave-vector  $k_{\mu}$ is not satisfying any massless dispersion relation.  Instead, the dispersion 
 relation is determined by the source, such as, e.g.,  
  cosmological constant. 
 
 Our postulate is that the above classical solution, quantum-mechanically is represented by a state vector in 
 some Fock space, built by a set of creation, $a^+_{I,k}, a^+_{L,k}$ and annihilation,
 $a_{I,k}, a_{L,k}$, operators, which satisfy the usual commutation relations, 
    \begin{equation}
   [a_{I,k}, a^+_{I,k'}] \, = \, \delta^4 (k - k')\,,~~\,   [a_{L,k} ,a^+_{L,k'}] \, = \, \delta^4 (k - k') \, 
   \label{commutator}
   \end{equation}
 with all other commutators vanishing.  Since the states created by these operators, correspond to states with some occupation number of longitudinal off-shell gravitons,  the 
  four-momenta $k$ do not satisfy any a priory dispersion relation. The relation among them 
  is determined by the state in which they are prepared. In other words, the information about $k$ is carried by the state vector. 
  
   The contact with the classical picture is then made via large-$N$ limit of expectation values, 
   \begin{equation}
 b_{k}^* \, = \,  {L_P\over \sqrt{\hbar}} \langle N + 1|  a^+_{k} |N\rangle \, |_{N\, \rightarrow \, \infty, ~ H=fixed} \, .
 \label{explimit}
\end{equation}
Correspondingly,  all subsidiary conditions must be understood to be satisfied on states and expectation values.

   A would-be classical weak-field de Sitter space in our picture is described as the quantum state 
 $|N_H\rangle $, with the following matrix elements, \footnote{We could have chosen instead 
 a coherent state description,
 \begin{equation}
 |deSitter\rangle \, = {\rm e}^{-{N \over 2}} \, \sum_0^{\infty} \, {N^{{n\over 2}} \over \sqrt{n!}}  |n\rangle \,, 
 \label{coherent}
 \end{equation}
 on which the expectation value of the destruction operator is non-vanishing
 $ \langle deSitter| \,  a \,  | deSitter\rangle \, = \, H\sqrt{\hbar} \sqrt{N} [...]$  and reproduces the classical metric
 in $N=\infty$ limit. However, the physics of the processes that we are going to discuss is better captured 
 in the number-eigenstate  representation.} 



\begin{eqnarray}
 \langle N_H + 1|  a^+_{I,k} |N_H\rangle  \, &= &\,
 \sqrt{\hbar} H \sqrt{N_H +1} \, [ \delta(k_0) \delta^3(\overrightarrow{k}) \, - \,  \\ \nonumber
 && \delta(k_0 - H) \delta^3(\overrightarrow{k}) ] 
 \label{matrixI}
\end{eqnarray}
  and 
 \begin{eqnarray}
 \langle N_H + 1|  a^+_{L,k} |N_H\rangle  \, &=& \, \sqrt{\hbar} H \sqrt{N_H +1} \,[ 2 \delta(k_0) \delta^3(\overrightarrow{k}) \, -\,  \\ \nonumber  
 &&\delta(k_0 - H) \delta^3(\overrightarrow{k})\,  -\,  \delta(k_0) \delta^3(\overrightarrow{k} \,  - \, \overrightarrow{n}H )] 
 \label{matrixL}
\end{eqnarray} 
  where $N_H \, = \, (HL_P)^{-2}$. 
  
    We are now ready to understand the quantum evolution of a probe particle interacting with the above graviton condensate.  Let the energy momentum of a quantum particle  $\phi$ be  
    $\tau_{\mu\nu}(\phi)$. 
   Propagation of a particle on a linearized classical metric perturbation on Minkowski vacuum is given by the coupling,      
  \begin{equation}
  \int d^4x \, \tau^{\mu\nu}(\phi) \, h_{\mu\nu}  \, .  
 \label{interactionlinear}
 \end{equation} 
 In our picture this propagation  should emerge as a large-$N$ limit of quantum transition 
from an initial 
state $|\phi_{in}, N_H\rangle \, = \, |\phi_{in}\rangle \times |N_H\rangle$,  where 
$|N_H\rangle$ describes the  quantum counterpart of the de Sitter state, to a final state in which the occupation 
number of gravitons with wave number $k=H$ changes by one,  
  $|\phi_{f}, N_H\pm1\rangle = |\phi_{f}\rangle \times |N_H \pm1\rangle$. Transition to this state corresponds to a process
in which a particle deposits (absorbs) a single graviton into (from) the condensate and evolves to a final state
$|\phi_{f}\rangle$.  This process is described by the first Feynman diagram  in Fig.1.
  \begin{figure}[ht]
\begin{center}
\includegraphics[width=85mm,angle=0.]{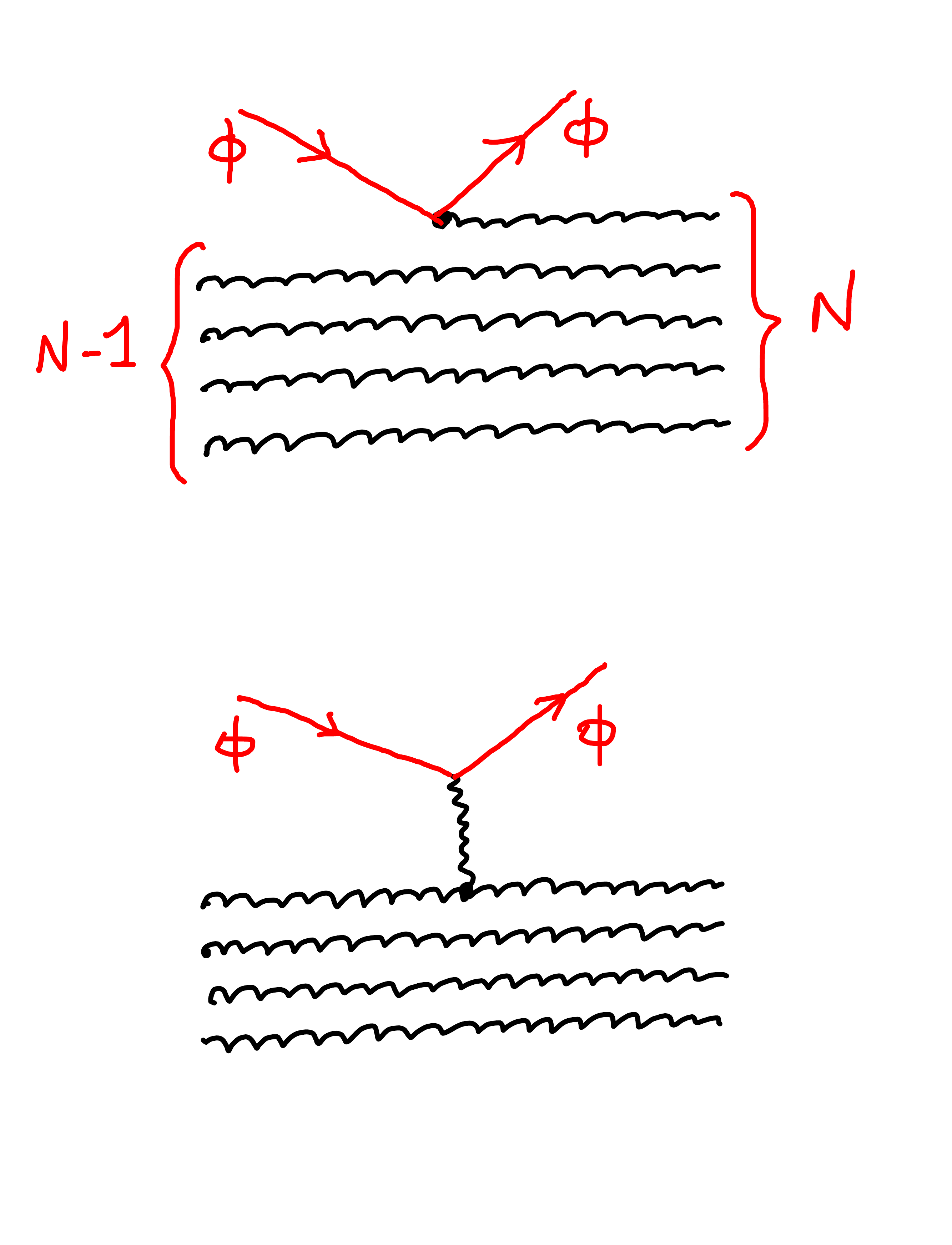}
\end{center}
\caption{Quantum scattering of a probe particle $\phi$ at the constituent gravitons.
}
\label{fig_safe}
\vspace{0.5cm}
\end{figure}

This is very similar to the process of  stimulated emission of photon by 
an excited atom. The emission probability is increased in the range of frequency corresponding to maximal occupation number of photons in the incident wave. 

 Such emission (absorbtion) of graviton are described by the following matrix elements 
 \begin{equation}
  \langle N_H + 1|  a^+_{k} |N_H\rangle \,,~~   \langle N_H - 1|  a_{k} |N_H\rangle \,.
 \label{aaa}
 \end{equation}
By taking into account (\ref{explimit}), (\ref{matrixI}) and (\ref{matrixL}), it is clear that 
to leading order in $tH$ and $rH$ expansions we recover the following effective classical metric,  
 \begin{equation}
 h_{ij}  \, \simeq \,L_PH \sqrt{N_H} 
 [I_{ij} t^2 \, + \, L_{ij} (t^2 \, + \, r^2) ] H^2 \, .  
  \label{recoverymetric}
 \end{equation}

 The crucial point is that because of the criticality of the graviton condensate, the prefactor
 is one , $L_PH \sqrt{N_H} \, \simeq 1$, up to $1/N$-corrections.  Thus, the above expression 
 recovers the classical de Sitter metric $h_{\mu\nu}$ given by  (\ref{linearDS}). Thus,  in the 
 large-$N$ limit, the graviton emission (absorption) transitions generate an $s$-matrix element 
 describing a scattering of a quantum particle $\phi$ in a {\it classical} de Sitter geometry with 
 all the usual consequences. 

Notice what has been the key point of the previous construction. The de Sitter condensate is defined as a particular collection of off-shell gravitons. The metric fluctuation satisfying the equation of motion in the presence of the cosmological constant source appears as the expectation value  on the condensate state i.e., as a collective phenomenon of the condensate itself. 

 In order to shed some more light  on the physical meaning of the longitudinal graviton condensate it is useful
 to visualize the condensate of off-shell gravitons in Einstein's  theory, as a condensate of on-shell gravitons 
 in the deformed theory, obtained by giving to gravitons a small mass, and then taking the zero-mass limit. 
 We can directly use the results of \cite{giajustinstefan}, where the 
  de Sitter solutions in such deformed theories were already analyzed. 
 
  It is useful to do this analysis in de Donder gauge, $\partial^{\mu} (h_{\mu\nu} \, - \, {1 \over 2} \eta_{\mu\nu} \, h) =0$. 
   In this gauge the equation  (\ref{cc}) takes the form 
\begin{equation}
 \square \left (h_{\mu\nu} \, - \, {1 \over 2} \eta_{\mu\nu} \, h \right ) \, = \, \eta_{\mu\nu} H^2 \,, 
 \label{gaugeDS}
 \end{equation}
 which has a de Sitter solution, 
   \begin{equation}
 h_{00} \, = \, -{1 \over 2} H^2t^2 \,, ~  h_{0i} \, = \, {1\over 3} H^2trn_i \,, ~~ \, h_{ij} \, = \, H^2 ({1\over 2} t^2\delta_{ij}\, + \, {1 \over 6} n_in_j|_{i\neq j}  r^2) \, .     
 \label{linearDSgauge} 
 \end{equation}  
  We now deform the theory by adding a mass term, so that (\ref{gaugeDS}) becomes, 
  \begin{equation}
 (\square\, - \, m^2) \left (h_{\mu\nu} \, - \, {1 \over 2} \eta_{\mu\nu} \, h \right ) \, = \, \eta_{\mu\nu} H^2 \, .
 \label{gaugeDSm}
 \end{equation}
The de Sitter solution of the massless theory is now deformed to 
 \begin{eqnarray}
  h_{00} &= &  \, -\,  {H^2 \over m^2}  (1 \, -\,  {\rm cos} (mt))\,,  \\ \nonumber
  h_{0i} &=& {H^2 \over 3m}\,  {\rm sin} (mt) \, n_i r \,, \\ \nonumber
   h_{ij} &= &  \, {H^2 \over m^2}  (1 \, -\,  {\rm cos} (mt)) \delta_{ij} \,  + \, {H^2 \over 6} {\rm cos}(mt) r^2 n_in_j|_{i\neq j} \,.
 \label{oscillator}
\end{eqnarray}
 
  This solution describes an oscillating massive bosonic field and quantum-mechanically represents   
a Bose-Einstein condensate of zero-momentum {\it on-shell} massive bosons.  At the same time for $tm \ll 1$, it reproduces 
the de Sitter solution (\ref{linearDSgauge}) of the massless theory.  Thus, the constituent gravitons that are 
off-shell from the point of view of the massless theory, are at the same time on-shell from the point of view of the massive one.  

The physics of this connection is the following. 
By introducing the mass term, the longitudinal gravitons of the massless theory became propagating degrees of freedom 
and got on shell \footnote{ Notice, that the free theory defined by equation   (\ref{gaugeDSm}), 
 $(\square\, - \, m^2) \left (h_{\mu\nu} \, - \, {1 \over 2} \eta_{\mu\nu} \, h \right ) \, = \, 0$, propagates six degrees of freedom, 
 one of them being a ghost (with the wrong sign of the kinetic term).  This can be seen in a number of equivalent ways. 
 For example, the above equation is the effective equation satisfied by the two transverse tensorial polarizations of the
 Pauli-Fierz massive graviton, obtained by integrating out the longitudinal polarizations,  and thus, viewed as a full theory it cannot be ghost-free\cite{gia-strong}.  Indeed, viewed as a full linear theory it is equivalent to a massive theory with a ``wrong" (non-Pauli-Fierz) mass term,  
$(h_{\mu\nu}h^{\mu\nu} \, - \, {1\over 2} h^2)$, which propagates ghost.   To see this equivalence, it is enough to notice 
 that in the latter theory a would-be gauge condition $\partial^{\mu} (h_{\mu\nu} \, - \, {1 \over 2} \eta_{\mu\nu} \, h) =0$ is a constraint that follows from the equation of motion as a result of Bianchi identity. 
Existence of the ghost in the above theory is however  unimportant for the present discussion, since we are considering 
only massless gravitons anyway.  We just wanted to stress that from the point of view of rendering  the longitudinal degrees of freedom physical,  the bound-state of gravitons has an effect remotely-analogous to the mass.  For non-linear completions of the 
theories of the type (\ref{gaugeDSm}) see \cite{maggiore}.}.
 However, they continued to form the same physical condensate as in the massless theory.  
In particular a probe particle  $\phi$ interacting with gravitons, is unable to distinguish 
the two cases for $tm \ll 1$.  

  The usefulness of the above discussion lies in showing that the off-shell longitudinal gravitons  are as legitimate 
  constituents of the condensate as the on-shell massive bosons.

 Coming back to the scattering of the probe $\phi$-particle, we can now summarize what is the underlying quantum picture of this evolution. 
    Take a process in which the initial particle $\phi$ deposits a single graviton of Hubble wave-length into the condensate and increases the occupation number exactly by one unit. 
     At the same time  $\phi$ is getting red-shifted 
    by conservation of energy and momentum (recall  in this picture the vacuum is Minkowski 
    so both quantities are conserved). 
  
%
%

%

 The fact that this should match the classical evolution, could have been guessed by the following 
 computation of the rate of the process described above.  The process is the lowest order in 
 $G_N$ and has an energy-transfer $H$. The gravitational  coupling is thus, $\alpha \, = \, (HL_P)^2$, since the graviton creation takes place on an $N$-graviton state, this gives enhancement factor 
 of order $N_H$ in the rate (equivalently of order $\sqrt{N_H}$ in the amplitude).  The rate of the process is thus, 
 \begin{equation}
  \Gamma \, = \, (\alpha N_H)  \, H \, = \, H \, . 
 \label{ratedeposit}
 \end{equation}
  Thus,  for the energy loss by the incident particle in time we get, 
  \begin{equation}
   {\dot{E} \over E } \, = \, - \, H \, , 
 \label{energyloss}
 \end{equation}
  which exactly matches the  rate of the energy loss  due to redshift on a classical expanding de Sitter   
 branch. 
 
   The lowest order non-linear correction to the classical metric can be recovered by  substituting 
   $h_{\mu\nu}$  with  ${L_P \over \sqrt{\hbar}} \hat{h}_{\mu\nu}$ in  
   (\ref{nonlinearexchange}) and taking the matrix element, 
       \begin{equation}
 \langle \phi_{f}, N_H|  \int d^4x d^4\tilde{x} \, \tau^{\mu\nu} (\phi(x)) \, \Delta_{\mu\nu, \alpha\beta} (x \, - \, \tilde{x}) \,T^{\mu\nu}(\hat{h}(\tilde{x}))  |\phi_{in}, N_H\rangle\, ,    
 \label{nonlinearelement}
 \end{equation}  
 This transition corresponds to a quantum process in which the $\phi$ particle scatters with one 
 of the constituent gravitons via one virtual graviton exchange, however without  knocking it out of the condensate. This process is described  by the second Feynman diagram in Fig.1.  
 The occupation number of gravitons is not changing in this process. 
  So the relevant graviton matrix element is now bilinear in creation and annihilation operators
  $\langle N_H|  a^+ a |N_H\rangle$ which amounts to a factor $N_H$. However,  this enhancement 
  is overcompensated  by an additional  suppression factor $H^2L_P^2 (Ht)^2$. So overall 
  the  (\ref{nonlinearelement}) is suppressed relative to the linear contribution by a factor 
  of order $(Ht)^2$  (or $(Hr)^2$).  This is not surprising, since in the 
  large-$N$ limit the matrix element (\ref{nonlinearelement}) reduces to a matrix element
  describing the scattering of $\phi$-quanta  
  in an effective classical metric
        \begin{equation}
 h_{\mu\nu} (x) \, = \, \int d^4\tilde{x} \Delta_{\mu\nu, \alpha\beta} (x \, - \, \tilde{x}) \,T^{\mu\nu}(h(\tilde{x})) \,,    
 \label{effectivemetric2}
 \end{equation}  
where $T^{\mu\nu}(h(\tilde{x}))$ is evaluated on $h_{\mu\nu}$ given by (\ref{linearDS}).  Obviously, this is nothing but the first nonlinear correction to the classical metric in the weak field expansion.  
  
   To summarize, we have outlined how the  quantum scattering at the constituent gravitons, in large-$N$ limit,
  imitates motion in the background classical metric.  
  
  \section{Particle Creation as Non-Vacuum Process}
  
   We now wish to discuss, the quantum process, which in the semi-classical limit 
   recovers the familiar particle creation process on a curved background, such as Hawking 
   evaporation.  In particular,  for the de Sitter case, we  shall recover Gibbons-Hawking \cite{deSitter} particle creation.    
  In the case of inflationary universe,  analogous process recovers creation of gravity waves 
 \cite{Inflation2} or curvature perturbations \cite{slava1, slava2}. 
 However, there exist a fundamental difference in the underlying physical phenomenon. In all the standard 
 cases \cite{deSitter, Inflation2, slava1, slava2}, the particle creation on a curved background is a vacuum process.  The crucial difference in our case is that no such vacuum processes are possible, since our vacuum is Minkowski with globally-defined time.  Instead particle creation in our case is the consequence of 
  {\it quantum depletion} of the actually existing particles of the condensate.  Only in a unphysical 
 limit of strictly infinite $N$,  the two descriptions become identical.  However, because 
 $N$ is always finite, for long time-scales our picture gives dramatically different result. 
 
  The leading contribution to a particle-creation process comes from two-in-two scattering, during which the two constituent gravitons re-scatter in such a way that one becomes pushed out of the condensate 
  and becomes propagating.    Such a process can be achieved either by a contact four-graviton vertex, or via one intermediate virtual graviton exchange.  
 
  Both processes contribute at the same order into the particle creation.  So for definiteness, let us discuss the second process. The corresponding Feynman diagram is given in Fig.2. 
    \begin{figure}[ht]
\begin{center}
\includegraphics[width=85mm,angle=0.]{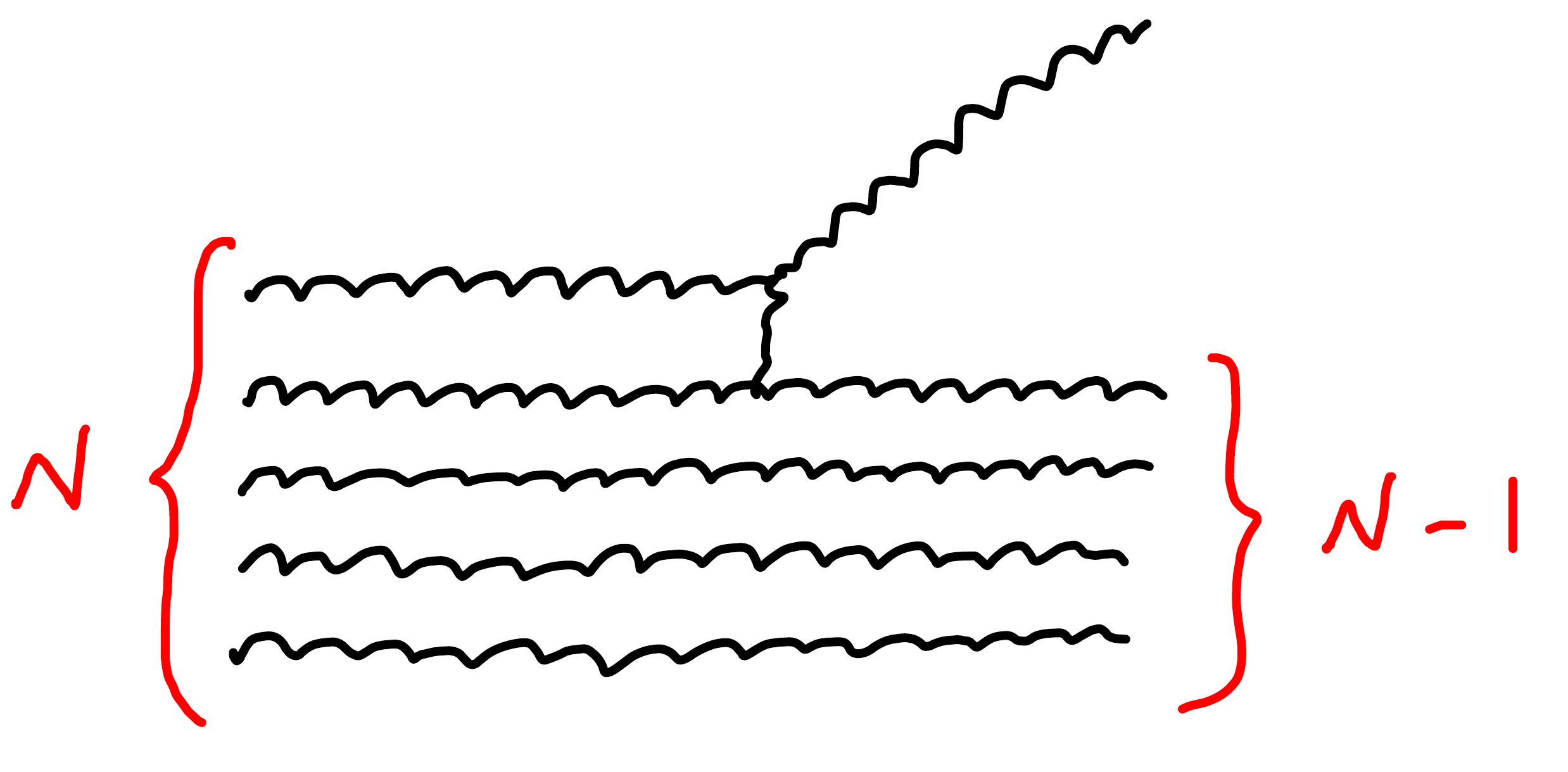}
\end{center}
\caption{Leading order process responsible for the particle creation. 
}
\label{fig_safe}
\vspace{0.5cm}
\end{figure}
  
  This transition is described by the following matrix element
         \begin{equation}
 \langle 1_{H+k}, 1_{H-k} N_H|  \int d^4x d^4\tilde{x} \, T^{\mu\nu} (\hat{h}(x)) \, \Delta_{\mu\nu, \alpha\beta} (x \, - \, \tilde{x}) \,T^{\mu\nu}(\hat{h}(\tilde{x}))  |N_H\rangle\, ,    
 \label{particlecreation}
 \end{equation}  
    where $k$ is the momentum transfer in this process.  For momentum transfer of order $H$, the above 
matrix element is of order  $\sqrt{N_H(N_H-1)} H^2 L_P^2 \, \sim \, 1$.  This is because the matrix element on two creation and two annihilation operators is 
          \begin{equation}
 \langle 1_{H+k}, 1_{H-k} N_H|a_{H+k}^+ a_{H-k}^+ a_H a_H |N_H\rangle\, \propto 
  \sqrt{N_H(N_H-1)} H^4/\hbar^2 \, .    
 \label{particlecreation1}
 \end{equation}  
The extra $L_P^2/\hbar$ factor is coming from the couplings, whereas one of the  $H^2/\hbar$ factors is absorbed by the propagator.  Thus, the depletion rate, is $\Gamma \, \sim \, H$. 
That is, the graviton condensate leaks one graviton of  Hubble wavelength per Hubble volume per Hubble time.  This is the physics behind the Gibbons-Hawking particle creation \footnote{Strictly speaking Gibbons-Hawking effect implies that a geodesic observer in de Sitter space will detect thermal radiation. This thermal radiation however does not imply a priori any form of evaporation. The main reason is because Gibbons Hawking thermality is not breaking any de Sitter symmetry. By contrast the effect we are describing in this section implies a net change of the value of $N_H$ by depletion. Of course this change will not lead to any form of evaporation in the $N=\infty$ limit with $H$ finite. This is in practice the limit where Gibbons-Hawking computation is done. }

 When the graviton condensate is not pure, but contains the admixture of a second Bose-gas,
 of some $\phi$-particles,  the new depletion channels open up, since now 
 gravitons can re-scatter at $\phi$-s. Correspondingly, if the occupation number of 
 $\phi$ quanta is larger than that of gravitons, the depletion  will be enhanced 
 by a factor $N_{\phi} \over N$.  As we shall see, this is the reason behind the 
 enhancement of the curvature inflationary density perturbations \cite{slava1} relative to 
 gravity waves\cite{Inflation2} or Gibbons-Hawking particle creation \cite{deSitter}. 
 
  We must comment on the following fact.   The  above picture also explains why in  
  classical static metrics with globally-defined Killing time the quantum creation of particles 
  does not happen.  In a condensate of longitudinal (and/or temporal)
 gravitons that in large-$N$ limit reproduces such classical static background metric, the particle creation by depletion does not take place. 
  This is for example true for the graviton condensate that corresponds to static AdS metric, 
   \begin{equation}
 h_{00} \, = \, - H^2 r^2\,,~~ h_{0i} \, = \, 0 \,,~~ \, h_{ij} \, = \, -\, H^2 n_in_j r^2 \,.   
 \label{linearADS} 
 \end{equation}  
The reason is that the constituent gravitons in such a case carry zero frequencies.  Thus, by energy conservation, their re-scattering cannot result into creation of positive-frequency particles.  In contrast, 
in time-dependent metrics this obstruction does not arise, since the constituent gravitons carry non-zero frequencies.  
 Notice, that absence of momenta in the constituent gravitons is not an obstacle for depletion, since 
 created particles can carry opposite momenta even if initial gravitons had zero momenta. 
   However, the final gravitons cannot carry the opposite energies, since in our picture the particle creation is not a vacuum process 
 and all the particles are real.

 In this respect, we should not be confused by the fact that the de Sitter also can be written in a static patch, which is obtained by taking the positive sign in front of $H^2$ in (\ref{linearADS}).  But,  this seemingly-innocent change makes a dramatic difference, since written in this gauge the space-time is no longer globally-defined and the extension beyond the horizon introduces the time-dependence.  
 The same applies to the black holes.   The constituent longitudinal gravitons corresponding to these two spaces carry non-zero frequencies and particle creation by depletion follows.  
  
  The above makes clear physical sense, since a static patch of de Sitter metric for $Hr \ll 1$ is very similar to 
  the metric created by an uniform-density static sphere of radius less than $H^{-1}$ near its center.  
  Such a sphere is neither a black hole nor de  Sitter, but the short wave-length particles 
  near the center cannot tell the difference from the de Sitter space. 
  
  Obviously, in such a classical metric no particle creation out of vacuum should take place. This semi-classical fact is explained in our picture by the fact that the 
 corresponding longitudinal gravitons cannot create particles due to lack of positive frequencies. 
    Contrary, a time-dependent metric leads to particle creation at arbitrarily short distances, 
    and this is explained in our picture by the possibility of depletion into positive energy modes.

 The fact that the particle creation is not a vacuum process, has far-reaching consequences for 
 nullifying the  black hole paradoxes as well as for predicting new properties, such as 
 existence of a measurable quantum hair \cite{Hair}. 
%
     Before  discussing implications for inflation and de Sitter, we would like to  confront the two philosophies of thinking about  particle-creation as vacuum versus non-vacuum  process. 
    
    \subsection{Compositeness and the Nature of Time}
    
 It is pretty obvious that in any quantum field theory defined on a space-time possessing a global time Killing vector, the vacuum 
pair-creation processes are forbidden. This is not the case when we quantize fields in space-times without a globally-defined time Killing vector. In those cases, since the notion of vacuum cannot be globally defined, we need to connect different local definitions by Bogoliubov transformations that eventually lead to real pair-creation. Typical, although different, examples are the de Sitter and the black hole metrics. This sort of vacuum pair-creation is a semi-classical phenomenon for which  the geometrical space-time arena in given and quantization is defined, accordingly, relative to such background geometry. Geometrical back reaction to these processes is extremely hard to define within this semi-classical frame. Indeed,  any back reaction modifies, among other things, the notion of time, and therefore, we should re-accommodate the rules of quantization in each step. 

An obvious approach to overcome these difficulties is to try to track this complicated dynamics using an auxiliary system with a well-defined global Killing time. This is in essence the goal of the holographic duals of gravitational systems, where the hope is to describe, in a dual field theory defined on a space-time with a globally-defined time,  the gravitational phenomena, such as black holes, where we lack (once we include the inner and outer space) a globally-defined Killing time vector. Actually, the touchstone of holography is to account for processes that semi-classically seem to be the consequence of the lack of a globally-defined Killing time (as, for instance, Hawking radiation) in terms of a theory (a so-called ``holographic dual") where time is associated with a global Killing vector. Our compositeness approach to gravity is an alternative attempt to address this apparent conundrum \footnote{For a recent discussion on this sort of puzzles see \cite{Pol} and references therein} from a new angle.

Indeed, as already stressed, the compositeness approach to quantum geometry relies on representing geometry in terms of condensates of off-shell gravitons. In this approach the quantization is defined relative to Minkowski geometry where among other things we count with a globally-defined time-like Killing vector and therefore with a well-defined notion of conserved energy. The only ingredient needed to define curved geometry is to work with condensates of off-shell gravitons that collectively behave as solutions to Einstein's equations. The notion of ``on-shellness"  is required not for the constituent gravitons, but for the collective modes of the condensate state, i.e., in a manner that is condensate-state-dependent.

In this approach quantization is defined relative to Minkowski space-time, while geometry is defined as collective phenomenon of the graviton condensate. The key advantage of this point of view is that we can use standard quantization rules as well as the key notions of energy and energy-conservation in order to account for the physics of curved geometry.

Within this frame an obvious question is to understand the graviton condensate counterpart of those classical geometries that lack a global Killing time.

The discussion for the concrete example of de Sitter space-time in the previous section already contains the clue to the answer. Indeed, the condensate counterpart of the lack of global Killing time is the spontaneous quantum depletion. As already stressed, this phenomenon only depends on the scattering processes among the constituent gravitons provided those condensates have non-vanishing number of off-shell gravitons possessing non-vanishing frequencies. The lack of global time for the corresponding classical geometry is determined by the semiclassical limit of the quantum depletion rate. 

The crucial point of this result is that what in the semiclassical approximation appears as
a consequence of the lack of a globally-defined time (i.e., the vacuum pair creation) in composite picture becomes simply a depletion due to scattering among constituent gravitons. This means that the entanglement of the created pair is not predetermined to be maximal by the geometry, but instead depends on the particular dynamics of the scattering processes underlying the quantum depletion. Note again that the pivotal ingredient lies in defining curved geometry as condensates of
off-shell quanta instead of defining on-shell quanta relative to a background curved geometry.

  \section{De Sitter and Inflationary Universe}

   In order to be able to interpolate between de Sitter and inflationary spaces
    we shall "regularize" de Sitter by replacing it with a finite lifetime 
  quasi-de Sitter cosmological space that asymptotically evolves towards Minkowski \footnote{This way to proceed will allow us to avoid the problems inherent to the lack of de Sitter $S$ matrix \cite{Smatrix1}, \cite{Smatrix2}, \cite{Smatrix3}}.  In this case, we can consistently define the notion of constituent gravitons and apply our reasoning.  In doing so we shall discover a remarkable fact.  {\it There is no consistent limit in which one can recover eternal de Sitter with well-defined classical metric descritipn}.  In the quantum picture in which background compositeness is taken into account only inflationary spaces of finite life-time can exist.  Any theory that in the classical 
  limit would flow to a state with a constant positive curvature is inconsistent.   Not surprisingly this observation has important implications both for cosmological constant problem, as well as for 
 post-inflationary measurements of cosmological parameters. 
   
  We shall now discuss how this picture comes about. 
  
 In order to regularize de Sitter  we are forced to introduce other type of quanta into the game.  Namely, we need 
   to introduce a cosmological fluid that plays the role of a homogeneous cosmic clock. This role is played by a scalar field, 
  the {\it inflaton}, which we shall denote by $\phi$.  Thus,  in our quantum picture we are forced to consider a state composed of 
   two types of quanta: gravitons and inflatons. In order to understand whether such a fluid can be at the quantum critical point let us study the system in more details. 
   
   Assuming the simplest form of the scalar potential $V(\phi) \, = \, {1 \over 2} (m/\hbar)^2 \phi^2$, the classical 
   evolution of the system with homogeneous scalar field is described by the usual set of equations
   \begin{equation}
   \ddot{\phi} \, + \, 3 \, H \, \dot{\phi} \, + \, \left ( {m \over \hbar} \right ) ^2\phi \, = \, 0 \, ,   
   \label{cosmic}
   \end{equation}
   and 
   \begin{equation}
   H^2 \, \equiv  \, \left ({\dot a  \over  a} \right )^2\, = \, {L_P^2 \over 2\hbar} \,   \left (({m \over \hbar})^2\phi^2 \, + \,\dot{\phi}^2 \right ) \, , 
   \label{hubble} 
   \end{equation}   
   where $a(t)$ is the scale factor and dots stand for time derivatives with respect to the cosmic time $t$. 
   For the time being we have kept the Planck constant $\hbar$ explicit. 
   Classically, it is well known \cite{Chaotic}  that this systems exhibits the two well-understood regimes. 
   
 For $\phi \,  \ll \, \sqrt{\hbar} /L_p$,  the friction term is  small, $H\, \ll \, {m \over \hbar}$,  and 
 the  scalar field undergoes damped oscillations.  In this regime the Universe expands as matter-dominated.  
  On the other hand, for  $\phi \, \gg \,  \sqrt{\hbar} /L_p$    
 the friction due to Hubble expansion is dominant $H\, \gg \, {m \over \hbar}$, and  the universe inflates. 
  The departure from the de Sitter state is thus measured by the slow-roll parameter $\epsilon \, \equiv \,  m^2/(\hbar H)^2$, which for the particular case of  $V=m^2\phi^2$ potential controls the number of e-foldings,  
  \begin{equation}
  {\mathcal  N}_{e}  \, = \, \epsilon^{-1}  \, .
  \label{Ne-folds}
  \end{equation}

  Our prescription is to think about this seemingly-classical system quantum mechanically as 
  the mixture of the two Bose-gases. Let us compute the occupation numbers in the two 
  regimes.  Consider the oscillating regime first. 
   The occupation number of gravitons per Hubble volume, $R_H^3 \, \equiv \, H^{-3}$,  is easy to estimate
 and is given by 
 \begin{equation}
 N \, = \, (R_H/L_P)^2 \,. 
 \label{numberg}
 \end{equation}
  On the other hand the number density of scalars is $n_{\phi} \, =\,  {m \over \hbar^2} \phi^2$. Thus, the occupation number per Hubble volume is  given by 
  \begin{equation}
 N_{\phi} \, =\, n_{\phi} R_H^{3} \, =\,  {\hbar H \over m} \left ({R_H \over L_P} \right )^2 \,. 
 \label{scalarN}
 \end{equation}
 Therefore we observe that the ratio of the occupation numbers, 
 \begin{equation}
   {N \over N_{\phi}} \, = \, {m \over (\hbar H)}  \, = \, \sqrt{\epsilon} \,, 
 \label{numbers}
 \end{equation}
  is measured by the slow-roll parameter.  
  Notice that since $\phi$-quanta attract gravitationally with the strength $\alpha_{\phi} \, = \, 
  (m \, L_P/\hbar)^2$, we have the relation, 
  \begin{equation}
  \alpha_{\phi} N_{\phi}  \, = \, \sqrt{\epsilon}  \,. 
  \label{critparameters} 
  \end{equation}
 Thus, the criticality parameter of the inflaton Bose-gas is the slow-roll parameter.  In the oscillation 
 phase we have $\alpha_{\phi} N_{\phi} \, \gg \, 1$.  
  Implying that, in the matter-dominated phase, the Bose-gas of scalars is over-critical 
and must be unstable. This is indeed the case. This is the well-known instability towards gravitational clamping of non-relativistic matter, which in the language of Bose-Einstein condensate amounts to instability towards breaking of translational invariance due to formation of  localized lumps (so-called bright solitons) that collapse.

 Let us now see what happens in the inflationary phase.  Classically, inflation takes place for $\epsilon \, < \, 1$. Thus, classically, the critical value $\epsilon \, = \, 1$ separates the inflationary regime from  matter 
  domination.  
 
  Quantum-mechanically this critical value has two very interesting meanings.

 First,  from eq(\ref{critparameters}) we observe that this value is 
  at the same time the critical point of {\it quantum}  
 phase transition for the inflaton Bose-gas to Bose-liquid. It is obvious that the system cannot maintain itself at this point for any significant time scale, both because of the redshift due to the expansion, as well as because of the quantum instability of the $\phi$-condensate towards clamping. 
   
  The second quantum meaning of $\epsilon \, = \, 1$ is that at this point the occupation numbers of inflatons and gravitons are equal. This is clear from eq(\ref{numbers}). 
 For   $\epsilon \, <  \, 1$, the inflaton occupation number starts to dominate and the system reacts by storing 
 the access of $N_{\phi}$ in form of the vacuum energy which drives inflation. 
 
    Generalizing the counting of the occupation numbers for arbitrary  potential $V(\phi)$, we  
 can write the classical time evolution equations for $N_{\phi}$ and $N$ in the following instructive 
 form
  \begin{equation}  
      \left({\dot{N} \over N} \right)_{class} \, = \, \epsilon \,  H \, ~~ {\rm and} ~~\, \left({\dot{N}_{\phi} \over N_{\phi}} \right)_{class} \, = \, \eta \,  H 
 \label{evolutionclass1}
 \end{equation}     
 where $\eta \, \equiv \, (V''M_P^2/V \hbar)$ is the second standard slow-roll parameter. 
 We see that understanding inflation in terms of occupation numbers, sheds a new light on the physical meaning of the slow-roll  parameters. As we see, these parameters control the rate of {\it classical} relative change  of graviton and inflaton occupation numbers respectively.  Inflation ends when either of the rates 
 becomes order one.

  Viewed from this quantum perspective inflation reveals a new identity.  What is usually seen as a classical 
 exponential expansion can be regarded as a quantum reaction 
 of the system whenever the inflaton occupation number dominates!
  As we shall see, the system also tries to restore  the balance, by  depleting the quanta stored in the background. It is this depletion that limits the duration of inflation.  
 
 In order to build-up confidence in the power of our picture we shall go step by step, by first showing that it reproduces all the known inflationary predictions as a particular approximation. Then, we shall 
 go beyond this approximation and discover effects that are not captured by the standard treatment.  
 
  \subsection{Quantum Origin of Inflationary Perturbations}  
  
   As a first consistency check of our quantum portrait we shall reproduce the 
   well-known inflationary predictions on the spectrum of perturbations. 
    In the standard picture of inflation the density perturbations is computed within the semi-classical treatment \cite{slava1},\cite{slava2}, \cite{Inflation2}. In this treatment the background is described by classical fields, $\phi(t)$ and $a(t)$, and one 
    considers small quantum perturbations on top of  this classical background.  Therefore, in this 
    treatment the quantum-compositeness of the background remains unresolved and all the corresponding 
    quantum effects are unseen.  As we shall see, this standard computation can be recovered as a special limit of our full quantum 
    picture.    
    
    In our approach, the  would-be classical background is resolved and treated as a Bose-Einstein condensate. This implies that
    what in the standard treatment appears as creation of small perturbations from the vacuum, in our 
    case  is represented by quantum depletion of the Bose-condensate.  During this depletion, some particles are pushed out of the condensate to occupy the higher excited quantum levels.  As explained above, the depletion is especially strong when the condensate is at the critical point. 
       Specifically for an observer that cannot resolve the compositeness of the condensate, the depletion looks as 
   quantum creation of particles out of the vacuum.  Of course, in this approximation,  we recover 
   the original semi-classical result.     
   
   In order to understand how our approach incorporates the standard picture as a particular limit it would be convenient to distinguish the following sequence of limits.

   $~~~$

    {\bf 1)  Classical limit. }  
    
     $~~~$
    
    Classical description corresponds to the  limit: 
    $\hbar \, = \, 0, L_P=0$,  whereas $G_N, m,\, =\, $ are kept fixed. In this limit the number of constituents $N = \infty$ whereas the Hubble radius stays  finite and the system becomes classical.  
   
   $~~~$
    
    {\bf 2) Semi-classical limit.}
    
    $~~~$
     
   Now $\hbar $ is kept non-zero, but  $N$ is still infinite at the expense of taking $L_P =0$ (i.e., 
   $G_N \, = \,  0$).  Notice that Hubble is still fixed to be finite at the expense of taking 
   energy density large, by $(m/\hbar)^2\phi^2 \rightarrow  \infty$.   In such a case, the background is still classical, but 
   the quantum fluctuations on top of it are permitted.  This is the limit in which the standard 
   analysis has been done.  
   
   $~~~$
   
    {\bf 3) Fully quantum picture.}
 
 $~~~$
    
     Finally, the fully quantum picture is achieved when all the parameters, $\hbar,  G_N$ and $m$  are kept finite.  In this case $N$ and $N_{\phi}$ are finite and  the background is no longer a classical field, but becomes a fully quantum entity with large but {\it finite} number of constituents. 

$~~~$

    It is very important to understand that  option 3) cannot be reproduced within 2) even if one takes into account all the higher order non-linear effects of perturbations. Such effects amount to the expansion 
    in powers of $\hbar$, but do not capture corrections in $1/N$, which as we shall see are extremely important. 
    
    Hence we shall start from 3) and recover 2) as an approximation in which background quantum compositeness 
   effects are neglected.  Next we shall compute the standard cosmological observables.

\subsection{Curvature Perturbations} 

  We shall first compute the inflationary density perturbations in our language. Within our framework 
 the source of density perturbations is quantum depletion of the condensate.  
 
  The background energy density rewritten in terms of graviton occupation number is given by 
  \begin{equation}
   \rho \, = \,  V \, =  {N \over R_H^3} \,  { \hbar \over R_H} 
 \label{energydensity}
 \end{equation}
 where $R_H$ is counted as the de Broglie  wave-length of constituent gravitons. 
 The density perturbation then is given by, 
   \begin{equation}
  \delta \rho \, = \, {\delta N_{\lambda}  \over R_H^3} \,  { \hbar \over \lambda} \,  
 \label{deltadensity}
 \end{equation}
 where $\delta N_{\lambda}$ is the number of the depleted particles (both inflatons and gravitons) of wave-length $\lambda$. 
  This quantum depletion takes place because the background gravitons and inflatons re-scatter 
  and are pushed out of the condensate.  Thus, in our picture, generation of the perturbations  
 is {\it not } a vacuum process in which the background only plays the  role of an external catalyzer of  
 particle-creation out of the vacuum. Instead, the created particles are actually emptying the background reservoir.  
   Since the gravitational interaction among the constituents is extremely weak ($\alpha \sim 1/N$), the 
   depletion is dominated by two-particle scatterings. 
 
  Thus,  the dominant contribution 
 is given by the re-scattering of  two constituent particles in the condensate during which one 
 gets pushed to a higher energy level.   Since the condensate has two components, there are three types of 
 possible scatterings: graviton-graviton, inflaton-inflaton and graviton-inflaton.  
 
 Notice that the de Broglie wave-length of the background gravitons is $\lambda_g \, = \,R_H$, whereas for inflatons it
is essentially infinite.  Because of this the re-scattering is dominated by the processes with graviton participation, in which the momentum transfer is $\sim \hbar H$ and correspondingly the 
effective gravitational coupling is $\alpha \, = \,  H^2L_P^2$.  The similar coupling 
in the inflaton-inflaton scattering is negligible. 

 Observe, that the gaviton-graviton and graviton-inflaton scattering rates only differ  by a multiplicative  combinatoric factor that is counting the number of available particle pairs in the two cases. Since $N_{\phi} \, \gg \, N$
 the dominant contribution into the depletion comes from the graviton-inflaton scattering, 
 \begin{equation}
    g \, + \, \phi  \rightarrow \, g \, + \, \phi  \, , 
\label{rocess} 
 \end{equation} 
 with the characteristic momentum transfer  $\sim \, \hbar H$.    The rate of such a depletion process is, 
 \begin{equation}
   \Gamma_{g\phi} \, = \,  H \, \alpha^2 N_{\phi} N \, , 
   \label{gphi}
   \end{equation}
   where $\alpha \, = \, (L_P H)^2$ and where the combinatoric factor $N_{\phi}N$ counts the number of graviton-inflaton pairs. 
Using the near-criticality relation, $\alpha N \, = \,1$, the above rate  can be rewritten as
\begin{equation}
   \Gamma_{g\phi} \, = \,  H \, {N_{\phi} \over  N} \, .  
   \label{gphi1}
   \end{equation}
  This depletion rate can be translated into the quantum evolution equation for the occupation numbers of gravitons and inflatons as, 
   \begin{equation}
   \dot{N} |_{quant}  \, = \, \dot{N}_{\phi} |_{quant} \, = \,- \, \Gamma_{g\phi} \, = \, - {1 \over \sqrt{N} L_P}  \, {N_{\phi} \over  N} \, .  
   \label{ratesquantum3}
   \end{equation}  
Notice the similarity with the depletion equation (\ref{leakage}) for a black hole. The difference lies in the enhancement 
factor ${N_{\phi} \over  N}$. Such enhancement factor for the black hole case counts the number of available light particle species, which increases  the number of depletion channels according to (\ref{leakagemany}). 
 Similarly, in case of inflation the large occupation number of inflaton quanta enhances the number 
 of depletion channels.  
 
   From (\ref{ratesquantum3}) it is clear that the number of depleted quanta per Hubble time with 
   Hubble wave-length is given by  $\delta N \,=  \delta N_{\phi} \, = \, {N_{\phi} \over  N}$, with the corresponding energy 
   density 
     \begin{equation}
  \delta \rho \, = \,  \hbar H \, {1 \over R_H^3} \, {N_{\phi} \over  N}  \, .  
 \label{deltadensityN}
 \end{equation}
   Notice, that since neither inflaton quanta nor the  longitudinal gravitons in the condensate carry tensor helicity, the above 
   expression contributes only to the scalar mode of density perturbations. In order to evaluate this contribution, we simply have to compute the Newtonian potential produced by the above energy 
   at Hubble scale. The answer is given by,   
\begin{equation}
\delta_{\Phi} \, = \, \delta \rho \, R_H^2L_P =  \,  L_P \, H \, {N_{\phi} \over  N} \,.  
   \label{deltascalar}
   \end{equation}
Expressing the classical values of the occupation numbers through  the slow roll parameter (\ref{numbers}) we recover the well-known expression for the density perturbations \cite{slavabook}, 
\begin{equation}
\delta_{\Phi} \, = \, L_P \, {H  \over \sqrt{\epsilon}} \,.  
   \label{deltascalarusual}
   \end{equation}

    The tensor mode of density perturbations 
 comes from the graviton depletion but this time due to graviton-graviton re-scattering and therefore the corresponding rate is suppressed relative to $\Gamma_{g\phi}$ by the factor  $N/N_{\phi}$. Thus, we have, 
 \begin{equation}
   \Gamma_{gg} \, = \,  H \,.   
   \label{gg}
   \end{equation} 
  Correspondingly, for the power-spectrum of the tensor modes we have,   
  \begin{equation}
\delta_{T} \,  =  \, L_P \, H \, .  
   \label{deltatensor}
   \end{equation}   
 Thus,  for the ratio of tensor to scalar perturbations we obtain, 
  \begin{equation}
  r \, \equiv \,   {\delta_T^2 \over \delta_{\Phi}^2} \, =   {N^2 \over N_{\phi}^2} \, = \epsilon\, .  
   \label{rparameter}
   \end{equation}

 Applying the above results for a particular case of  $m^2\phi^2$ inflation, we can express
 ${N_{\phi}^2 \over  N^2}$ in terms of the number of e-foldings  via 
 (\ref{Ne-folds}). In these conditions equations (\ref{deltascalar}), (\ref{deltatensor}) and 
(\ref{rparameter}) turn into the  well-known semi-classical expressions for cosmological observables in terms of the number of e-foldings. Namely, 
\begin{equation}
 \delta_{\Phi} \, = \, L_P \, H \,  \sqrt{{\mathcal N}_e}\, ,   
   \label{deltascalar1}
   \end{equation}
 and
 \begin{equation}
  r \, = \,  {1 \over {\mathcal N}_e} \, .  
   \label{rparameter1}
   \end{equation}
   
 \subsection{Tilt}  
   
   The tilt can be computed according to the standard definition, $n_s \, - \, 1\, = \, {d ln \delta_{\Phi}^2 \over d ln k}$, 
  which has to be evaluated at the point of horizon crossing. This definition for our case translates 
  as $n_s \, - \, 1\, = \, H^{-1} \Gamma^{-1} {d  \Gamma \over d t}$.  In order to evaluate this quantity 
  let us rewrite  the rate (\ref{gphi1}) in terms of occupation numbers according to 
  (\ref{ratesquantum3}), 
\begin{equation}
   \Gamma \, = \, {N_{\phi} \over N^{3/2} L_P} \, .
\label{gammainN}
\end{equation}
Taking the time derivative, we get 
\begin{equation}
1 -n_s \, = \, -  {\dot{\Gamma} \over H\Gamma} \, = \, 
     {3\over 2}  \left({\dot{N} \over N} \right) \, \, - \left({\dot{N}_{\phi} \over N_{\phi}} \right) \, .   
\label{tiltinN}
\end{equation}
Accordingly, if we take into account only the purely  classical evolution of the occupation numbers through 
(\ref{evolutionclass1}) and ignore their quantum depletion (\ref{ratesquantum3}), we immediately recover the standard result,
   \begin{equation}
   1\, - n_s \,= \,  {3 \over 2} \epsilon \, - \, \eta \,.  
   \label{tiltclassic}
   \end{equation}
  In particular, for $V=m^2\phi^2$, this gives
  \begin{equation}
   1\, - n_s \,\simeq \,  {2 \over {\mathcal N}_e} \,.  
   \label{tilt}
   \end{equation}
  Consequently we recover all the standard results as a particular approximation of our quantum picture. 
 In this computation, although we have taken into account the quantum contribution of depletion
 as the source of particle excitations above the background, nevertheless, we have ignored the effect of this 
 depletion on the background itself. In other words,  
  we have ignored the finiteness of the reservoir, treating it as an infinite capacity source of particles.  
 Thus, we effectively worked in the limit of $N \, = \, \infty$, but keeping the ratio $N_{\phi}/N$ 
 (equivalently ${\mathcal N}_e$) finite.  
   Next we shall take into account finite $N$ corrections and show that these corrections reveal  something extremely important about inflation.   
  
\subsection{Quantum Corrections} 

  The key new ingredient that our picture brings into inflation is an additional {\it quantum} clock originating  from the compositeness of the background.   
   In the standard semi-classical treatment of inflation there is a single, {\it classical},  clock, set by 
   a slow-rolling scalar field, which leads to a classical decrease of $H$. 
  Of course, this clock continues to be present in our case and corresponds to 
  $\hbar \, = \, 0$ limit of our picture.  In our full quantum treatment, this is the clock that 
  increases $N$  and $N_{\phi}$ according to (\ref{evolutionclass1}), due to the classical time-evolution of $H$. 
   Correspondingly, this classical clock is also responsible for producing the tilt in the spectrum due to the classical decrease of the momentum transfer in the depletion process and also the decrease of $N_{\phi} / N$ ratio.  
  All these effects have the semi-classical counterparts, as we have demonstrated by explicitly recovering 
  semi-classical observables in the previous section. 
  
     Novelty in our case is the existence of the second, {\it quantum}, clock. These clock works differently 
     from the classical one, and unlike the latter, decreases both $N$ and $N_{\phi}$ according to  
     (\ref{ratesquantum3}) 
     due to quantum depletion of the background condensate.  This is something unheard off in the conventional treatment, 
     since in semiclassical approach the production of perturbations is a vacuum process and 
    there is no notion of background depletion. For us the story is very different, since our background 
    is a finite reservoir of gravitons and inflatons subjected to depletion. The  most interesting thing 
    is that the depletion of the background is a cumulative effect and thus encodes information about the entire history 
    of the Universe. That is,  the quantum state of an inflationary patch at some ${\mathcal N}_e$, depends on the number of prior e-foldings, $\Delta {\mathcal N}_e$, since the beginning of inflation!
    
     We can identify at least two distinct sources through which the prior expansion history is encoded in
     this quantum state of the background.  One source is entanglement, which will be discussed later
     and the second one is the depletion of the background gravitons.    
     By taking into account  these two effects,  an observer can scan the entire history of the 
     inflationary patch, way beyond the last $60$ e-foldings.       
     
      In this section we shall focus on the effect of quantum depletion. 
      The story is nicely summarized by the following master equations
    \begin{equation}
  {\dot{N} \over N}  \, = \, H \left ( \epsilon \, -  \,  {1 \over \sqrt{\epsilon} } {1 \over N} \right ) \, , 
   \label{master11}
   \end{equation}
    and 
  \begin{equation}
  {\dot{N}_{\phi} \over N_{\phi}}  \, = \, H \left ( \eta \, -  \, {1 \over N} \right ) \, , 
   \label{master22}
   \end{equation} 
   describing the time evolution of the occupation number of the background gravitons and inflatons respectively. 
   The first terms of the r.h.s. of both equations stands for the classical evolutions due to slow-roll. These 
  evolution increases the Hubble 
   radius and thus the occupation number of gravitons. The occupation number of the inflaton field per Hubble
   volume also increases, but  slower than gravitons, so that the graviton occupation number catches up.  
  These classical-evolution terms are of course non-zero in the limit
   $\hbar \, = \,  0$.   In this limit the equations (\ref{master11}) and (\ref{master22}) evolve 
   into  (\ref{evolutionclass1}). 
      
   The second terms in the r.h.s. of equations (\ref{master11}) and (\ref{master22}) come from the quantum depletion and have  the  opposite 
   effect.  These terms vanish in the classical limit, since in this limit $N\, = \, \infty$.  This equation, already gives the first warning, that for finite number $N$, the number of e-fodings cannot be arbitrarily large without  running into inconsistency.   Indeed, it is inconsistent to make $\epsilon$ arbitrarily small, since eventually the depletion rate will blow up.

    This equation shows that the depletion of gravitons and inflatons is a cumulative effect.
   For example, for  $V= m^2\phi^2$ inflation 
  the decrease of the occupation number in the background during $\Delta {\mathcal N}_e$ number of e-folding since the onset of inflation is given by, 
  \begin{equation}
  \Delta  N\, = \, \Delta  N_{\phi} \, = \, - \Delta {\mathcal N}_e^{{3\over 2}}  \, .
   \label{changeofN}
  \end{equation} 
  This immediately implies the upper consistency bound on the total number of  e-foldings 
  as,  
  \begin{equation}
    {\mathcal N}_{Total}^{{3\over 2}}  \,  < \, N.
   \label{boundonchange}
  \end{equation} 
Notice that since the depletion is a cumulative effect over many e-foldings, the bounds cannot be removed by any re-summation.  It is a consistency bound and not an artifact of some perturbative 
treatment.  In particular it implies that eternal de Sitter limit $\epsilon=0$ is inconsistent. We shall come back to this implication, but in the meantime let us discuss the effect on cosmological observables assuming that the bound is satisfied. 
  
   We can distinguish two types of new quantum contributions. 
   
   The first one arises 
   due to additional functional dependence of $N$ and $N_{\phi}$ on ${\mathcal N}_e$ coming from the 
   second term in the r.h.s. of (\ref{master11}) and (\ref{master22}), which is absent in the classical treatment. This dependence 
   gives an additional, quantum, contribution into the tilt, 
   \begin{equation}
    \Delta (1 \, - \, n_s)_{quantum} \, = \, - \,  {1 \over N}  {N_{\phi} \over N}  \, = \, - {1 \over \sqrt{\epsilon}  N} \, ,    
    \label{deltatilt}
    \end{equation}
  By   taking into account in (\ref{tiltinN}) 
    the full quantum time dependence of $N_{\phi}$ and $N$ via 
 (\ref{master11}) and (\ref{master22}) we obtain the quantum-corrected tilt 
 \begin{equation}
   1\, - n_s \,= \,  {3 \over 2} \epsilon \, - \, \eta \,  - \, {1 \over \sqrt{\epsilon} N}   \, . 
   \label{tiltfull}
   \end{equation}

 Using a concrete form of the scalar potential, the above correction can be translated in terms of 
 the number of e-foldings.  For example, for  $m^2\phi^2$ inflation this gives, 
    \begin{equation}
     (1 \, - \, n_s)_{total} \, = \, {1 \over {\mathcal N}_e} \, - \, {\sqrt{{\mathcal N}_e} \over N} \, .     
    \label{deltatilt}
    \end{equation}
   This correction into the tilt is local in time and it is measured by the remaining number of 
   e-foldings towards the end of inflation.  There are few remarkable things about it. 
   First, unlike the standard contribution given by the first term, it is more important at the earlier times. 
    Secondly, it  reinforces the upper bound  (\ref{boundonchange}) on the number of e-foldings, since 
  it starts to  exceeds the semi-classical contribution as soon as the bound is violated.  Again it excludes the possibility of eternal inflation \cite{selfreproduction 1}, \cite{selfreproduction 2}, \cite{selfreproduction 3} ,\cite{selfreproduction 4},\cite{selfreproduction 5},\cite{selfreproduction 6},\cite{selfreproduction 7} or any scalar potential classically permitting a constant plateau region.  
  
  The value of the inflationary Hubble parameter favored by the cosmological measurements, $\hbar H \, \sim 10^{13}$ GeV,  corresponds to 
  $N \, = \, 10^{12}$, which makes the above correction negligible for last $60$ e-foldings. Nevertheless, it has a fundamental value  as it restrict the number of e-foldings and eliminates the eternal de Sitter as a consistent limit of a slow-roll inflaton potential as well as eternal inflation.   
  
    Perhaps from the observational perspective more interesting is the second type of contribution 
  that has a cumulative nature and thus can have a stronger measurable effect  provided  the total number of e-foldings is large. 
  This contribution affects the cosmological observables, due to the depletion of  $N$ and $N_{\phi}$ accumulated throughout the inflationary history.  The cumulative quantum change of occupation numbers
  between some initial and final time moments is  given by integrating  the second terms
  in the equations (\ref{master11}) and (\ref{master22}) and is given by,    
  \begin{equation}
  \Delta  N\, = \, \Delta  N_{\phi} \, = \, -  \int_{t_{in}}^{t_f} \, {H \over \sqrt{\epsilon} } \, dt \, .
   \label{changeintegral}
  \end{equation} 

   Given the explicit form of the inflationary potential $V(\phi)$ this change can be expressed 
   as the function of e-foldings since the beginning of inflation.  
  For example, for $m^2\phi^2$ inflation  the 
  occupation numbers of gravitons and inflatons  after  $ \Delta {\mathcal N}_e$ e-foldings since the beginning of 
  inflation changes according to the equation (\ref{changeofN}).   
    In other words, the would-be de Sitter background 
 contains less gravitons and inflatons  as compared to the beginning of inflation,  even if the classical evolution  of the Hubble is frozen.

   The change in the occupation numbers (\ref{changeintegral}) must be taken into account when computing the cosmological parameters.      
  In particular, taking into account (\ref{rparameter}) the accumulated quantum change 
 of $r$ parameter (up to terms suppressed by $N/N_{\phi}$) is given by 
   \begin{equation}
   \Delta r\, = \,   \ 2 {N  \over  N_{\phi}^2} \Delta N \, ,  
 \label{incrementR}
 \end{equation} 
 where $\Delta N$ is given by (\ref{changeintegral}). 
  The corrected value of $r$ is thus, 
   \begin{equation}
   r\, = \,  \ {N^2 \over  N_{\phi}^2} \left ( 1 \, - \,   {\Delta N  \over N} \right ) \, . 
 \label{changer1}
 \end{equation} 
 Applying this result to a particular case of  $m^2\phi^2$ inflation we get
  \begin{equation}
   r\, = \,  \ {1 \over  {\mathcal N}_e} \left ( 1 \, - \,   {\Delta {\mathcal N}_e^{{3\over 2}}  \over N} \right ) \, . 
 \label{changer1}
 \end{equation} 
  Notice that  ${\mathcal N}_e$ is the remaining number of e-foldings at the moment when the
  $r$-parameter is evaluated, whereas $\Delta {\mathcal N}_e$ is the number of e-foldings since the 
  beginning of inflation till that moment.  So the entire duration of inflation would be  the sum
  of the two numbers, ${\mathcal N}_{Total} \, = \, {\mathcal N}_e \, + \, \Delta {\mathcal N}_e$.
  
   The expression (\ref{changer1})  is remarkable in two respects. First it reinforces the same upper bound (\ref{boundonchange}) on the number  of e-foldings. Secondly, if this number is large, it gives a remarkable possibility of probing pre-$60$ e-folding history by the 
post-inflationary measurements of cosmological parameters.  For example, an inflation with $10^{7-8}$ e-foldings will give order-one contribution to $\Delta r/r$.   This is impossible in the standard treatment 
of inflation where background depletion is not taken into account.

 \section{Non Eternity versus Quantum Eternity} 
 
 One very  important conclusion that emerges from our picture is that for the finite values of Planck and Newton constants,   $\hbar$ and $G_N$, there exist no possible consistent  choice of parameters that could reproduce eternal de Sitter as a limit of slow-roll inflation. 
 
  In this part of the paper we would like to understand the physical meaning of this observation 
  and its possible connection to the cosmological constant problem. 
   
   What we are learning is that in a quantum world with gravity number of e-foldings is finite and very much limited. In short , {\it  we are discovering that a de Sitter Universe (defined as metric entity) as a limiting case of slow-roll inflation is inconsistent with quantum mechanics.}

    This discovery may sound very surprising if one judged from the standard (semi)classical intuition.    Indeed, classically,  we can make a scalar potential arbitrarily close to a constant. For example, 
    for $V(\phi) \, = \, \omega^2\phi^2$ (where $\omega \equiv m/\hbar$)  there is nothing wrong in taking the limit 
  \begin{equation}
  \omega \, = \, 0, ~ \phi \, = \, \infty \, ,
  \label{limitclass} 
  \end{equation}
   while at the same time keeping energy density $V(\phi)$ finite. 
   Classically,  such a field becomes frozen and the system evolves as an eternal de Sitter.  
       Is there any incompatibility with our findings?  Of course not.  In the classical theory, 
       since $\hbar \, = \, 0$,  the occupation number of  gravitons $N$ becomes infinite and number of e-foldings can be arbitrarily large. Thus, a {\it classical} eternal de Sitter 
       is fully compatible with our bound (\ref{boundonchange}).  
       
     Similarly, we can consistently obtain de Sitter as a limit of the slow-roll in the semi-classical 
   approximation.  In this case, we have to keep $\hbar \, = \, $fixed, but take 
   $L_P\, = \, 0$ (equivalently $M_P \, = \,\infty$) at the expense of
   $G_{N} \, =\, 0$.  At the same time the Hubble radius has to be kept finite and constant.  
  In familiar $\hbar \, = \, 1$ units, semi-classical de Sitter limit corresponds to the choice, 
  \begin{equation}
     H \, = \, {\rm fixed}, \, ~M_P\, \rightarrow \, \infty\,,~{\rm and} ~  \phi \, \gg \, M_P \left({M_P\over H}\right )^{{1 \over 3}} \,.  
     \label{DSlimit}
    \end{equation}
   It is obvious that with such a choice  $N\,  \rightarrow \, \infty$ and $\epsilon \, \rightarrow \, 0$, 
  in such a way that  $\epsilon \, \gg \, N^{-2/3}$, so that our limit on slow-roll (\ref{boundepsilon}) is always satisfied.

       However, Nature is not semi-classical but quantum and both $\hbar$ and $L_P$ are non-zero. 
   Consequently, $N$ is finite and the bound (\ref{boundepsilon}) prevents us from ever reaching 
   de Sitter as a consistent limit of slow-roll. 
     We shall now discuss the physical meaning  of this bound. 
     
  \subsection{Physical  Meaning of Non-Eternity Bound}  
     
      The inflationary system is characterized 
     by a set of parameters.   This set includes classical entities, such as, $\phi(t)$ and $a(t)$ (or equivalently $H$, $\epsilon$ and $\eta$) as well as the quantum ones, such as, $N$ and $N_{\phi}$. 
     We have established certain relations  between the quantum and classical parameters through 
     the quantum constants of Nature, such as, $\hbar$ and $L_P$.    Given the classical characteristics, 
  these relations enable us to read-off the quantum ones through the equations  (\ref{NN}) and 
  (\ref{epsilon}) and subsequently  derive their time-evolution (\ref{evolutionclass1}) in terms of the classical  parameters.  But, since the quantum description is more fundamental than the classical one, inevitably there are situations when the classical description in terms of  $H$, $\epsilon$ or $\eta$ stops to make sense and the only possible description is  in terms of quantum entities, such as, $N$ and $N_{\phi}$.  Our framework enables us to quantify this breakdown of the semi-classical description in terms  of the bound (\ref{boundepsilon}).  The remarkable thing about this bound is that it is telling 
  us that the breakdown of semi-classical description has nothing to do with trans-Planckian energy densities, but rather it comes in from the macroscopic life-time of the system.
  
    An inflationary system can be treated as approximately semi-classical if  the time 
 evolution of the  occupation numbers $N$ and $N_{\phi}$ can be reliably  deduced through their dependence on classical characteristics. That is, as long as equation (\ref{evolutionclass1}) is a good description.  In other words the quantum term in equations 
 (\ref{master11}) and (\ref{master22}) must be sub-dominant.  
 
  Therefore, the consistency of the description requires that over the time-scale of classical change 
  ${\delta H \over H} \, \sim \, 1$, the quantum contribution to the change must be still negligible.
  In particular, we must have  $\Delta N \, |_{quant} \, \ll \,  \Delta N \, |_{class}$ for the quantum and classical changes derived from equations (\ref{ratesquantum3}) and  (\ref{evolutionclass1}) respectively. 
 If this is not satisfied,  the system becomes intrinsically quantum and the description in terms of classical entities makes no sense.  This is the case for any inflationary potential that allows 
 $\epsilon$ to violate the bound (\ref{boundepsilon}).  In such a case, any region of the classical potential
 violating this bound is excluded. This conclusion has important consequences. In particular, it excludes the regime where self-reproduction or eternal inflation could take place within the validity of approximate semi-classical description in terms of the metric.  However, this does not exclude some new notion of {\it quantum eternity},  to be considered later.

 \subsubsection{Non-Eternity Versus Self-Reproduction} 
 
Notice that the bound (\ref{boundepsilon}) becomes saturated before one can reach
a so-called self-reproduction \cite{selfreproduction 1},  regime.  This  regime in our  quantum language would take place 
for 
\begin{equation}
  \epsilon_{self.rec.} \, = \, {1 \over N} \, , 
  \label{selfrec}
  \end{equation}
but this is excluded by our bound.   In the above regime, the evolution is entirely dominated 
by the quantum depletion ({\ref{ratesquantum3}).  In the would-be self-reproduction regime, the number of depleted quanta per one Hubble time is 
$\sqrt{N}$, which means that such an Universe, viewed as a classical space time, can only survive for 
$\sim \, \sqrt{N}$ e-foldings.  This is inconsistent with the very idea of self-reproduction. 
According to this idea, the quantum fluctuations play the role in pushing the inflaton field up  
the potential (or at least stopping its classical slide-down), but the classical geometric description  of the background space time is 
assumed to be still valid. 

  It is useful to see were the standard semi-classical reasoning about self-reproduction
 becomes incompatible with the quantum treatment.  
 
 The idea of self-reproduction is as follows \cite{selfreproduction1}.  Let us continue to parameterize inflation by  classical entities, such as $\phi$ and $H$, but  do not ignore the role of the quantum fluctuations for changing these entities. 
  In these conditions, the classical inflaton field $\phi$ is subject to time-dependence because of the following two sources. 
   The first source of time-dependence is the classical  slow-roll,
   \begin{equation}
    \dot{\phi} \, = \, - \, {V' \over 3H} \, .
\label{slow}
\end{equation}
 During one Hubble time, $t_H \, =\, H^{-1}$, this slow roll causes the decrease of $\phi$ by an amount, 
   \begin{equation}
    \delta {\phi}_{class}  \, = \, - \, {V' \over 3H^2} \,.
\label{slow}
\end{equation}
 On top of this classical time-evolution, we shall super-impose the quantum fluctuations.  These fluctuations 
 are assumed to cause random jumps of $\phi$ on the Hubble scale per Hubble time given by,
     \begin{equation}
    \delta {\phi}_{semi-class}  \, \sim \, H \, .
\label{jump}
\end{equation}
Demanding that the above two contributions cancel each other in some domains, 
   \begin{equation}
    \delta {\phi}_{class}  +  \, \delta {\phi}_{semi-class}  = \, 0\, , 
\label{canselation}
\end{equation}
we obtain the the following condition on the classical parameters,  
\begin{equation}
     {V' \over H^2} \, \sim H\,. 
\label{condition4}
\end{equation}
 Essentially, this condition is equivalent to demanding that the scalar mode of curvature perturbations 
 (\ref{deltascalarusual}) be order one, 
\begin{equation}
     \delta_{\Phi}  \, \sim 1\,. 
\label{largeperturbation}
\end{equation} 
 For example, in $V = m^2\phi^2$-inflation the above conditions imply,
 \begin{equation}
  \phi \, = \, M_P\sqrt{{M_P \over m}} \, .
  \label{phivalue}
  \end{equation} 
 It is then assumed that in such domains the field $\phi$ will be prevented from rolling 
 down the slope and the domain shall continue to inflate without decreasing the Hubble rate. 
  Thus, in this approach it is assumed that the role of quantum mechanics 
  is limited to being a force that counteracts the classical slow roll of $\phi$ but otherwise the space-time is assumed  to be well-characterized by classical entities, such as horizon. 
  
   Let us now show, that for finite $N$ the above interpretation is inconsistent.  Not surprisingly, this already follows from our bound
   (\ref{boundepsilon}), but we wish to understand what the correct quantum picture is. 
      In order to see this, let us translate 
   what would be the required values  of $N$ and $N_{\phi}$ for self-reproduction.   Translating 
   equations (\ref{condition4}) and (\ref{largeperturbation})  in terms of occupation numbers we get
   the relation (\ref{selfrec}). Since this value violates the bound (\ref{boundepsilon}), the evolution is 
   fully dominated by the quantum depletion (\ref{ratesquantum3}). Thus, we have, 
   \begin{equation}
     \dot{N}   \, =  \, H \sqrt{N}\,,  
\label{dotnself}
\end{equation} 
which means that within the Hubble time the quantum depletion reduces $N$ by 
  \begin{equation}
     \delta {N} _{quant}   \, =  \,\sqrt{N}\,.  
\label{deltaNself}
\end{equation} 
 Let us now evaluate the corresponding change of Hubble, assuming that it still makes sense to 
 talk about the classical horizon.  Then  taking into account 
 (\ref{deltaNself}) in relation  (\ref{NN}), we get  for the change of Hubble, 
  \begin{equation}
     \delta {H}_{quant}   \, =  \, H^2 L_P\,.  
\label{changeH}
\end{equation} 
 We have to compare this change, to the would be change of Hubble induced by the semi-classical jump
 of $\phi$.  The latter change is at most, 
  \begin{equation}
     \delta {H}_{semi-class}   \, \lesssim  \, mH L_P\,.  
\label{changeH}
\end{equation} 
Taking the ratio we get 
  \begin{equation}
    {\delta {H}_{quant}  \over \delta {H}_{semi-class}} \, \gtrsim \, {H \over m} \, \gg \, 1\,.    
\label{nonsense}
\end{equation} 
 This shows that the semi-classical argument leading to self-reproduction neglects a huge impact
 on $H$.  By taking this impact into account, we see that the space-time that enters the self-reproduction regime after ${\mathcal N}_e \sim M_P/H$ e-foldings looses any classical meaning.   In particular,  after this time 
 there is no possibility to characterize the Universe with a well-defined classical  Hubble radius.  
 
 As a final remark let us just mention that the above quantum difficulties on the possibility of eternal inflation can have some relevant consequences on the very idea of using eternal inflation as the mechanism to fill up the string landscape as well as on the notion of landscape itself.

  \subsection{Backgrounds Without Classical Analog?}
  
  The previous argument on self-reproduction was based on looking for a region in classical parameter space where the quantum fluctuations of the inflaton field -- once they are superimposed on the classical change in a Hubble time -- lead to the existence of domains of Hubble size where effectively the classical value of the inflaton field has not changed. In a nutshell the problem with this form of self-reproduction is that it requires values of $\epsilon$ where quantum depletion effects are dominant. However, we can develop a similar self-reproducing argument once we have resolved the background into constituents, namely looking for the regime where the quantum fluctuations of $N$ once they are super-imposed to the classical evolution of $N$ in a Hubble time, lead to the existence of domains where the average value of $N$ does not change.
  
  The solution to this exercise is obvious from the master equation (\ref{master11}) governing the evolution of $N$.  Namely, $\epsilon$ should saturate the bound (\ref{boundepsilon}),   $\epsilon \, = \,N^{-2/3}$. In these conditions we get $\dot N=0$. We could interpret this result as indicating that for this extreme value of the 
  slow-roll parameter we can find, after a Hubble time, some domains where $N$ does not change.  This value of $\epsilon$ is much larger than the would-be self-reproduction threshold $\epsilon \, = \, 1/N$ obtained from the semi-classical reasoning.   Thus,  if this reasoning were still valid, we were to conclude that  
despite the fact that in such domains $N$ is constant, the change of the classical value of the inflaton field is by no means vanishing. 
    
  The conflict now is similar to the one we have described above,  but this time the quantity that is not changing is, instead of the classical value of $\phi$, the value of $N$ and what creates the conflict is the semiclassical result of the change of the value of the inflaton field.  Indeed, we cannot keep $N$ constant, while allowing
 $\phi$ to evolve, without sacrificing the relation (\ref{NN}) between $N$  and the classical quantities.  
   
 The essence of this inconsistency is very clear.   
  In the case of the semi-classical self-reproduction argument we are thinking on how quantum effects affect the motion of a ball ($\phi$)  in a ``ready made" external background potential $V(\phi)$. However, 
  in the case of $N$-portrait we are directly addressing the quantum effects of the background itself. In this quantum picture there is no fixed external background. So $N$ is by no means "rolling" in any external potential and therefore $\dot N=0$ should be interpreted as defining a particular critical point of the background condensate itself. The mismatch between $\dot N=0$ and the semiclassical variation of the classical inflaton field simply indicates that we cannot interpret this state of the background condensate as representing the "ball" frozen somewhere in a classical pre-existing potential. In other words, the graviton background when $\epsilon$ saturates the bound (\ref{boundepsilon}) does not admit any semiclassical analog.

 \section{Implications for Cosmological Constant}

     In our picture a Universe filled with any form of potential energy represents a quantum entity composed out of 
     two ingredients. These are: 1) The quanta of the inflaton field  that make up the potential energy;  and  2) The gravitons 
     that are sourced by this energy.  The point is that the occupation number of gravitons, $N$, 
     is determined by the energy of the source, whereas 
     the occupation number $N_{\phi}$ depends on the time-dependence of its own energy density and becomes
     infinite in the limit of frozen energy density.  Correspondingly,  the depletion rate $\Gamma_{g\phi}$ of gravitons and
     $\phi$-s blows up in this limit. 
     
      The key difference of our approach with respect to the standard semi-classical treatment of inflation lies in relating the  quantum fluctuations to the sub-structure of the background, instead of interpreting them as the result of a vacuum process.  
In our picture the background is a {\it finite}  reservoir of gravitons and quantum perturbations are 
the result of their depletion.  Such a background is impossible to keep constant unless $N$ is infinite.  
 Only in $N=\infty$  case the semi-classical treatment becomes valid for an unlimited time. 
 
  Once we accept the finiteness of the background reservoir,  the physical meaning of the bound (\ref{boundonchange}) becomes very clear. 
  The bound simply comes from the consistency requirement that the condensate can survive the  quantum depletion during ${\mathcal N}_e$ 
e-folds.

   Thus, we are observing that  in any slow-roll inflation there is an  inevitable conflict between the classical and the quantum clocks.  The classical clock is trying to fill up the reservoir, whereas the quantum clock depletes it.  Whenever we try to make the classical clock to run slower than a certain critical rate, the quantum clock speeds up and the reservoir  is emptied before the classical clock has any chance to refill it.  The system becomes inconsistent. 
   
    As we have seen the speed-up of the quantum clock is caused by the increase of the
    inflaton occupation number due to slow-down of the classical clock.  This is inevitable whenever 
   the classical potential energy density that drives inflation can be resolved into constituents. 
   This in particular means that de Sitter cannot be reached as a consistent limit of slow-roll inflation. 
    
  But, what about a "dead" cosmological constant, which is not obviously resolvable  
 into quantum constituents? 
 
 The clue for understanding the difficulties underlying the notion of dead cosmological constant is to recognize that such a notion is normally defined using two potentially-incompatible descriptions, one classical and another intrinsically-quantum-mechanical. The classical description on which geometry is based, is given in terms of the Hubble constant $H$ with $\dot H\, =\, 0$, while the quantum description is given in terms of 
 number $N \, = \, R_H^2/L_P^2$ with $\dot N\, =\, 0$. The two relevant quantities $\dot H$ and $\dot N$ have completely different origin. Indeed, the one setting the classical clock, $\dot H$,  is evolving according to classical general relativity, while the one governing the quantum clock, $\dot N$, fully encodes quantum gravity effects as well as a concrete identification of what degrees of freedom the quantity $N$ actually counts. 
 
 In many approaches to the cosmological constant problem it is customary to identify $N$ with the de Sitter entropy\cite{deSitter} and in that sense to set the physical Hilbert space of the system as finite-dimensional \cite{Banks}. In those approaches the question about the nature of the $N$ quantum degrees of freedom as well as their dynamics is left unanswered. The approach we have pursued in this note simply identifies these degrees of freedom with soft gravitons of typical wave-length $R_H$. This identification has 
 far-reaching consequences.  In particular, it means that these degrees of freedom interact gravitationally and therefore the available set of  quantum states defines a standard infinite Hilbert space.  Only a sub-set 
 of states corresponding to a small nearly-gapless Bogoliubov collective excitations of the critical condensate 
 spans a finite dimensional portion of the Hilbert space. 
 
 In this approach the dynamics of the graviton system is governed by graviton-graviton scattering, which among other things induces depletion and sets the change of $N$ in time. Consequently,  getting $\dot N=0$ requires turning-off this interaction, i.e.,  sending $L_P$ to zero.  But, this is consistent with having a finite value of $H$ only in the $N=\infty$ limit. Therefore if we want to keep both $N$ and $L_P$ finite and at the same time to have $\dot N=0$ we need to give up the classical description in terms of $H$ with $\dot H=0$. The general lesson we learn from this discussion is that the classical characterization of cosmological constant as $\dot H=0$ appears to be inconsistent with keeping $N$ finite and $\dot N=0$. It looks that the classical and quantum characterizations of a dead cosmological constant are mutually inconsistent, or in other words {\it the quantum compositeness of gravity is incompatible with the classical definition of positive cosmological constant}. 
 
   We would like to comment that we are not the first to notice some potential issues with the de Sitter space. 
  The authors of \cite{Tsam} and \cite{Polya}, within the semi-classical treatment,  have observed
   effects that can be interpreted as instabilities of the de Sitter space.  Although, the precise connection  
 with our findings is not fully clear, both results point in the same direction.  The difference is that our framework, being 
 microscopic, has the potential of addressing the questions raised in\cite{Tsam, Polya}.  What we observe is that 
 de Sitter cannot be eternal in the ordinarily-used sense of a space that admits a classical metric description for
 an unlimited time.  We see precisely what goes wrong here. De Sitter, within a finite time, evolves 
 into an entity that looses half of its constituents and the remaining half presumably become maximally-entangled 
(see the next section).  Such a state can no longer be described classically, even approximately. 
 However, this does not prove that it is either inconsistent or non-eternal. 
 At the moment there is no evidence that such a system cannot last longer in a state of quantum eternity.

  What is the implication of this finding for the cosmological constant problem? 
  The entire story hangs upon a possibility of defining such a quantum state consistently.  
 If this can be done, then the cosmological constant problem is essentially unperturbed. 
  In order to illustrate this, let us assume that the curvature radius of an initially-classical de Sitter state 
  is one cm.  The initial occupation number of gravitons in such a state is $N \sim 10^{66}$ and the system 
  depletes one graviton per $10^{-10}$ sec.  In order to evolve into a non-metric state, the system would require the time-scale of order $10^{56}$sec.  Irrespectively what happens after, this time scale is so long that 
 the cosmological constant problem is essentially the same. 
  
  On the other hand, if the consistent quantum state with half-emptied de Sitter cannot be defined, the story will 
  change dramatically. In this case,  irrespective of the required time-scale for reaching such a state,  the positive  cosmological term could be rejected  {\it by consistency}.

 \section{Entanglement as Quantum Measure of de Sitter's Age} 

 We now wish to discuss one more phenomenon that makes quantum treatment of de Sitter and inflationary backgrounds 
 qualitatively different from the standard semi-classical approach. 
 In the latter case, since the composition of the background is unresolved,  
 the only information about the expansion history is carried by the semi-classical 
   perturbations on top of the background.  This information is very limited. In particular, an observer can measure only the imprints of last $50-60$ e-foldings and all the pre-existing history is erased 
   by inflation.  For example, it is impossible to detect the total number of e-foldings. 
   
    The situation in our quantum picture is fundamentally-different, because of the existence of  quantum clocks.     
     Unlike the classical case,  the background is a composite quantum entity and the information about the expansion history is imprinted in its quantum state.  In particular,  the history is encoded in form of the depletion and the entanglement of the constituent gravitons (and inflatons).  Thus, what is usually regarded as a classical background, in reality 
    is an entangled multi-particle quantum system.  
        This information about the generated entanglement is carried by depleted constituents. 
   Thus, by measuring the entanglement of the depleted quanta  at some moment of time 
 a hypothetical observer could in principle measure the absolute age of de Sitter. This is something        
fundamentally-impossible in the standard treatment of inflation. 

  Since within our picture both black holes and inflationary spaces are represented as critical graviton condensates, we  expect that they share close similarities also in efficiency of generating entanglement during their time-evolution.   This gradually-increasing entanglement  then acts as a second 
  quantum clock, which on one hand allows to scan the inflationary history and on the other hand 
  makes sure that classical description breaks down within a finite time scale. 
  
  The generation of the entanglement  in the composite picture of  a black hole was studied
  in \cite{P6, scramble}, using the prototype models.
 According to this study, the underlying reasons behind the generation of entanglement are:
 
\begin{itemize}
 
\item Near quantum criticality of the condensate
\item Quantum instability with respect to depletion
 
 \end{itemize}
 
  The roles of these items in generating entanglement can be described as follows.  The proximity to the quantum critical point ensures a  huge density of available micro states into which the system can evolve. These micro states originate from the fact that near the critical point order $N$ Bogoliubov modes of the condensate
 are crowded within $1/(NL_P)$ energy gap.   
 
  The instability due to quantum depletion triggers the effect of a "quantum roulette" facilitating the 
  exploration of these large density of states.  The characteristic time scale of  instability is set by the 
  Liapunov exponent, which is determined by the depletion rate $\Gamma$ of the graviton condensate.  
  
   In  \cite{scrample}  by analyzing a simple prototype it was shown  that the quantum break time 
 for the critical condensate is given by, 
    \begin{equation}
          t_{quant}  \, = \, \Gamma^{-1}  {\rm ln} (N) \, .
          \label{quantumbreak}
          \end{equation}
    This quantum break time sets the minimal time-scale for generation of entanglement.  The logarithmic dependence nicely matches a so-called fast-scrambling conjecture \cite{Scrambling2}, according to 
    which the black holes  should scramble information within a time that depends logarithmically on the entropy. 
   
   Notice, that the composite picture of black holes allows to deduce some new properties of the scrambling 
 time.  In particular the dependence on the number of light particle species, $N_{species}$, existing in the theory.       
  Since  the black hole depletion rate (\ref{leakagemany}) is sensitive to the number of extra light particle species,  so must be the time $t_{quant}$.  Thus, for black holes, the enhanced depletion rate should effectively shorten the entanglement (i.e., the scrambling) time, 
    \begin{equation}
 t_{BH}  \, = \, {L_P\sqrt{N} \over \, N_{species}}  {\rm ln} (N) \,  = \, {R_{BH} \over N_{species}}  {\rm ln} (N) \, .
          \label{quantumbreakBH}
          \end{equation}
          
 Analogous reasoning applies to de Sitter and inflationary spaces, since they satisfy 
both conditions necessary for fast entanglement. The graviton condensate is at the critical point and is unstable with respect to depletion.   
 
 Notice, that the classical expansion of the scale factor $a(t)\, \propto \, e^{Ht} $ should not be counted as an instability of the 
 condensate.  This time evolution only redshifts the probe particles that are excited above the background, but not the condensate itself.  In fact, in our picture the classical exponential redshift  of the probe particles 
 is a result of their quantum scattering with the graviton condensate.   The only instability of the condensate 
 that must be counted in pure de Sitter is the quantum depletion. 
 Hence,  we expect that pure de Sitter  must be also subjected 
 to a quantum entanglement clock, with the characteristic quantum breaking time given by 
    \begin{equation}
 t_{dS}  \, = \, L_P\sqrt{N}  {\rm ln} (N) \,  = \, R_{H} \,   {\rm ln} (N) \, .
          \label{quantumbreakds}
          \end{equation}
 This quantum clock of entanglement works against eternity of de Sitter.   After finite time, the de Sitter is no longer 
 describable  as a well-defined classical background. 
          
   Of course,  the same arguments must apply to inflation, except there is an enhancement in depletion due to 
 excess of inflaton quanta. As a result the quantum break time is shortened by the slow-roll parameter, 
     \begin{equation}
 t_{inf}  \, = \, \sqrt{\epsilon} \, L_P\sqrt{N}  {\rm ln} (N) \,  = \, \sqrt{\epsilon} \, R_{H} \,   {\rm ln} (N) \, .
          \label{quantumbreakInf}
          \end{equation}
 What we are discovering is that the entanglement clock, just as depletion of $N$, works against eternity of  
 inflation. 
 
  The two important time scales of the graviton condensate  are:  
  1) The time-scale during which the system becomes 
  one-particle entangled, $t_{one-p. ent}$; and  2) the time-scale of maximal entanglement, 
  $t_{max~ent}$.     
  
   In order to gain a better understanding on the entanglement phenomenon, let us think of an inflationary 
 universe as a graviton/inflaton condensate  starting in some initial  state which is well described by a classical metric. 
 Within our quantum picture, this means that the initial $N$-particle state in mean-field approximation is well-described by a single-particle wave-function and thus, is non-entangled.   
   Subsequently, the near-critical graviton condensate 
 evolves in two ways:

 1)  Undergoes depletion loosing its constituent gravitons (and inflatons);  

 2) Generates entanglement.

 We wish to focus here on the second process. 
  As a convenient unit of measurement we can take a one-graviton emission time, which also 
 defines the Liapunov exponent, and thus, quantum instability time. 
  This time-scale is set by the depletion rate (\ref{gphi1}), which in terms of slow parameter is $\Gamma^{-1} \, = \, \sqrt{\epsilon} \, R_H$.

The two time scales emerge. The first is the time during which the system becomes at least 
one-particle entangled. That is, if the initial $N$-particle state was representable (in some appropriate basis) as a  $N$-particle tensor product, 
\begin{equation}
\psi_1\times\psi_2\, ....\,\times\psi_N \, ,
\label{tensor}
\end{equation}
after $t_{one-p.ent}$ it can no longer be represented in the form $\psi_j\times \psi'$ for any $j$, where $\psi'$ is an arbitrary $N-1$-particle state. 
However, it could still be represented as some tensor product of multi-particle states.   
  This time scale $t_{one-p.ent}$ cannot exceed the the quantum breaking time (\ref{quantumbreakInf}).

   The second time-scale, $t_{max.~ent.}$ is the time after which the representation in form of a tensor product 
   is simply impossible and system becomes maximally entangled. 
  This time-scale is given by order $N$  depletion steps, $t_{max.~ent.} \, = \, N \, \Gamma^{-1}$. 
    That is, by the number of steps required to deplete abut a half of the graviton reservoir.  For the black hole case, this time plays the role of  Page's time \cite{Page}.  
 
  These time-scales can be visualized by the following group-theoretic parameterization introduced 
  in \cite{group}, 
 by labeling the black hole (in the present case de Sitter) state by a spinor irrep of $SO(2N+1)$ group. 
  Notice, that from the finite dimensionality of the spinor irrep by no means follows that the Hilbert space 
  of de Sitter in our picture is finite.   This parameterization only applies to a small portion of the Hilbert space 
  that accounts for nearly-degenerate Bogoliubov levels located within $1/N$ mass gap above the 
  condensate.   As we have stressed above, the entire Hilbert space of the system is of course infinite.

    We must also stress that the entanglement properties of graviton condensate are independent of this 
 group-theoretic description.  A skeptical reader can simply view this representation as an useful bookkeeping tool which allows to visualize the depletion and creation of entanglement in group-theoretic  
 terms.

   In this map, the  Cartan sub-algebra generators,  $\sigma_j, ~(j=1,2,...N)$,  correspond to the occupation number operators  of Bogoliubov modes, and their eigenvalues $\epsilon_j$ can correspondingly assume two possible values, $0$ or $1$.  Then an initial non-entangled state of de Sitter can be represented  by a basis 
   vector of spinor irrep of $SO(2N+1)$ that can be labeled by a set of $\epsilon_j$-s. 
   This can be visualized as $N$-long sequence of zeros and ones, 
   \begin{equation}
   |{\rm non-entangled~deSitter} \rangle \, \equiv    |\epsilon_1,\epsilon_2\, ...\epsilon_N \rangle \, .
   \label{initialN}
   \end{equation}
     
    Then, during one elementary depletion step, the sequence gets shortened by one unit i.e., 
   is mapped onto a spinor irrep of $SO(2(N-1)+1)$.  During this map it is no longer represented 
   by a single basic vector but by a superposition of minimum two basic vectors  that differ by 
   one random eigenvalue.  Thus, after every depletion step, the number of basic vectors in 
   the superposition roughly doubles.  Thus, after $m$ steps the initial state will evolve into 
   a superposition of minimum $2^m$ basic vectors of $SO(2(N-m)-1)$ spinor irrep, creating entanglement. 
   
    This group theoretic picture accounts that after $m_{one-P} \, = $ln$(N)$ steps the system should become minimum 
    one particle entangled, whereas the maximal number of steps for creating the total entanglement 
   must be $m_{total}  \,  \leqslant N/2$. 
    
      Thus,  expressed in terms of the Hubble parameter, the time-scale for which the inflationary 
      Hubble patch  becomes at least one particle entangled is (\ref{quantumbreakds}) 
    and (\ref{quantumbreakInf}) for inflation or pure de Sitter patches respectively.       
      Notice, that for the realistic values of the inflationary Hubble,  $H \, \sim \, 10^{13}$GeV, the minimal 
 entanglement time corresponds to roughly  $30$ e-foldings. Thus, if the total duration of inflation 
 is less than $80-90$ e-foldings, a hypothetical de Sitter observer  must be able to measure 
 an increase in entanglement among the depleted quanta  within  last $60$ e-foldings. 
 
 It is extremely important that in our quantum picture, generation of density perturbations is not a vacuum 
 process, but rather a depletion of {\it physically-existent} gravitons and inflatons of the condensate.    
 Because of this crucial difference, in our case, these depleted quanta carry information about the 
 entanglement of the background. 
        
   To be more concrete, let us assume that the total duration of inflation is ${\mathcal N}_e \, = \, 70$. 
  Then, the quanta created during the first few e-foldings will exhibit none or very little  entanglement.         
 Whereas, the subsequent quanta become more and more entangled reaching the  full one particle 
 entanglement  after $30$ e-foldings.  The emerging conclusions are:

  {\it  Observing increase of entanglement within last ${\mathcal N}_{ent}$ e-foldings would imply that 
  the total number of e-foldings is at least ${\mathcal N}_{Total} \, = \, {\mathcal N}_{ent} \, + \, 30$. }

   In particular, the created quanta cannot stay unentangled within the entire  last $60$ e-folds.    
    Also, observing the maximal entanglement from the beginning  of last  $60$ e-foldings would imply that total duration of inflation is larger than $90$ e-folds.  
    
    Of course, in this discussion we are only addressing matter-of-principle questions, without  touching 
    the observational issues of detecting entanglement imprints in post-inflationary measurements of the density fluctuation 
    spectrum.  But, this matter-of-principle points indicate fundamental difference between our and 
    standard semi-classical treatments, since they indicate that inflation is past-transparent for the 
    physical observations within a single Hubble patch.

 \section{AdS as Graviton Condensate}
The compositeness approach can be also extended to the case of negative cosmological constant i.e to AdS space time. The recipe is to represent these spaces as a graviton condensate with $R_{AdS}$ fixing the typical wave length as well as the occupation number of the gravitons defining the condensate. More specifically we shall consider 
 a homogeneous condensate of gravitons of wave length $R_{AdS}$ and occupation number -- per unit of space of size $R_{AdS}$ --  equal to
\begin{equation}\label{one}
N_{AdS} \, = \, \left({R_{AdS} \over L_P} \right)^{D-2}
\end{equation}
for $D$ the space-time dimension.

One of the main points of the graviton condensate model is to map classical geometry into a quantum state characterized by the graviton occupation numbers. In order to see how this works in the particular case of AdS let us consider a region of space of size $r$ and let us compute the number of gravitons inside this region for the AdS condensate defined by (\ref{one}). For the case of four dimensions we get,
\begin{equation}
N_{AdS}(r) \, = \, (r/R_{Ads})^3N_{AdS}\,  =\, r^3/(R_{AdS} L_P^2) \, .
\end{equation}
The first thing that this counting reveals is the well-known confining gravitational potential of AdS. Indeed the mean field potential created by these gravitons is simply given by,
\begin{equation}
V(r) \, \equiv  \, N_{AdS}(r) (L_P^2/r) \, = \, (r/R_{AdS})^2 \,,
\end{equation}
which agrees with the confining gravitational potential in classical AdS.

As we did in the case of black holes, we can define in the mean field approximation the effective coupling,
\begin{equation}
\lambda_{AdS}(r) \,  \equiv \, N_{AdS}(r) \, \alpha_{AdS}
\end{equation}
with $\alpha_{AdS}$ the coupling strength between two of the constituent gravitons, i.e., 
$\alpha_{AdS} \, = \, (L_P/R_{AdS})^2$. The criticality condition in this mean field approximation is determined by,
\begin{equation}
\lambda_{AdS}(r) \, = \, 1\,,
\end{equation}
that leads to $r=R_{AdS}$. This means that the AdS condensate reduced to a cell of space of size $R_{AdS}$ is critical and therefore -- according to our previous characterization of holography -- can be fully described in terms of the gapless Bogoliubov modes. In this sense it is interesting to notice \cite{Nportrait}  that the number of gravitons per cell of space at criticality indeed agrees with the central extension of the CFT in the standard AdS/CFT correspondence.  On the other hand, at the critical point the number $N_{AdS}$ sets the number of nearly gapless Bogoliubov modes of the graviton condensate.  Because of criticality, the physics of these 
Bogoliubov modes is conformal (at least up to $1/N_{AdS}$).  We are thus, observing an interesting 
fact, namely that
{\it  when viewed quantum-mechanically,   AdS space is a critical graviton condensate 
with number of nearly-conformal Bogoliubov modes $N_{AdS}$ equal to the 
central charge of CFT according to AdS/CFT conjecture.}

 It is natural to think that this equality is not just an extraordinary coincidence, and that the 
 two CFT's are actually one and the same.  An opposite conclusion would be hard to swallow,  since 
  would imply that we are discovering a second independent CFT with the central charge 
 that miraculously coincides wit the one suggested by the AdS/CFT correspondence. 
  Although, we cannot disprove such an option we consider it highly unlikely. 
   We are thus lead to the conclusion that the quantum foundation of AdS holography, just as in the case 
   of the black hole holography, is in quantum-criticality of the graviton Bose-gas, which makes the physics 
   of collective excitations approximately conformal. 
   
   The criticality of the condensate representing AdS can be easily understood in the spirit of the Wilson-Kadanoff approach to renormalization group. Indeed, if we consider the condensate at scale $r$, larger than $R_{AdS}$, and we try to describe it using gravitons of wave-length $r$, we shall need (in order to preserve the same amount of energy) $\frac{r^4}{R^2L_P^2}$ gravitons. This is exactly the same counting we did at scale $R$ i.e., the holographic counting $r^2/L_P^2$ corresponding to maximal packing, provided we scale the area of the boundary by $r^2/R^2$ (that is, once we introduce the scale factor defining AdS geometry). In other words, AdS geometry implements the invariance of the graviton condensate under Wilson-Kadanoff renormalization group transformations in the sense that at arbitrary scale $r$ the condensate always looks, for the AdS geometry, holographic, i.e., as composed of $ r^2/L_P^2$ gravitons of wave length $r$ . In this sense we can say that AdS geometry is the manifestation of the criticality of the graviton condensate.
   
   As a simple consistency check of this idea, we shall show that it correctly describes the physics of small and large black holes in AdS. 

\subsection{Black Holes in AdS}

In order to move on let us now add to AdS a gravitational field produced by some mass $M$ localized in a region of size $r$,  without affecting the AdS boundary conditions. 
 
 Again,  we shall treat this case in terms of the graviton occupation numbers. Adding a mass $M$ localized in a region of size $r$ is equivalent to adding a certain amount of gravitons of typical wave length $r$. More specifically 
\begin{equation}\label{three}
N_{M} \, = \, M^2L_P^2 \, , 
\end{equation}
gravitons of wave length $r$. Let us now look more carefully into the corresponding graviton system. It is composed of $N_{M}$ gravitons of wave length $r$ and of $N_{AdS}(r)$ gravitons of wave length $R_{AdS}$. For $r$ larger than $R_{AdS}$ the gravitons sourced by the AdS cosmological constant are harder than the ones sourced by the external mass $M$.  

Again, according to the condensate portrait of black holes, we shall characterize a AdS black hole of mass M by the value of $r$ that makes the previously-defined combined system of gravitons critical.

Before describing the criticality condition we can, already at this qualitative level, understand some main features of black holes in AdS. In the black hole portrait of asymptotically flat black holes, evaporation is understood in terms of quantum depletion. More specifically, the order of magnitude of the black hole temperature is determined by the escape energy of the constituent gravitons. Let us apply this logic to our AdS condensate. The escape energy for a soft graviton sourced by $M$, i.e., of wave length $r$, can be easily estimated to be
\begin{equation}
\epsilon(r) \,  \sim  \, N_{AdS}(r) (L_P^2/R_{AdS}r) \, = \, r/R_{AdS}^2 \, .
\end{equation}
Note,  that what dominates in this expression is the dragging created by the hard gravitons sourced by the cosmological constant. If we identify this escape energy with the expected temperature, i.e., with the typical energy of depleted gravitons, we observe that it scales with $r$, i.e.,  the system has effectively positive specific heat. Indeed the previous expression for the escape energy of soft gravitons agrees with the Hawking temperature of large black holes in AdS \cite{HP}.

Using this condensate model we can even figure out how soft gravitons ( we are considering the case 
$r\, \gg \, R_{AdS}$ ) can acquire enough energy to escape. The dominant effect corresponds to scattering processes of soft gravitons with hard gravitons of wave length $R_{AdS}$ (notice the similarity with the depletion process we encountered in case of inflationary universe due to re-scattering of soft inflaton quanta at hard gravitons).  

 This scattering is dominated with processes of momentum-transfers of the order of $1/R_{AdS}$. In average a soft graviton of wave length $r$ overlaps with $r/R$ hard gravitons if they are distributed homogeneously. Thus the net effect of hard gravitons is to change the wave length $r$ of the soft gravitons into a wave length $R^2/r$ corresponding to $r/R$ scattering processes with $1/R_{AdS}$ transfer momentum. In other words, the effect of the scattering with the hard gravitons is to blueshift the soft gravitons sourced by the mass $M$. Hence, in the regime of $r>>R_{AdS}$, we expect that the soft gravitons sourced by $M$ are blueshifted and consequently that its total number is reduced. To estimate its number after blueshifting we introduce the mean field effective coupling $\lambda_{M}(r)$ for the soft gravitons as we did for the hard ones sourced by the cosmological constant, namely
\begin{equation}
\lambda_{M}(r) \equiv N_M \alpha(r) \, , 
\end{equation}
with $\alpha(r) \, = \, L_P^2/r^2$ the strength of gravitational coupling among soft gravitons. After blueshifting we change,
\begin{equation}
\alpha(r) \rightarrow (r/R_{AdS})^4 \alpha(r) \, .
\end{equation}
Keeping the value of $\lambda_{M}(r)$ constant allows us to estimate the number of soft gravitons after blueshifting, namely: 
\begin{equation}\label{two}
N(r) \equiv (R_{AdS}/r)^4N_M \, .
\end{equation}
This is indeed a very interesting number whose meaning will become clear in a moment. Before we need to study the criticality conditions for the condensate in the regime $r\, \gg \, R_{AdS}$. The corresponding critical point will define the large black hole in AdS. 

In this regime the criticality condition will be defined by imposing that the typical energy of the gravitons sourced by $M$ is equal to the escape energy. In other words the system achieves criticality when the soft gravitons of wave-length $r$ are blue shifted by scattering with the hard ones and their number is reduced according to  (\ref{two}). This criticality condition immediately translates into the following self-consistent relation between $r$ and $M$:
\begin{equation}
M \, = \,  (R_{AdS}/r)^4N_M  \, (r/R_{AdS}^2) \, 
\end{equation}
that, after plugging the value of $N_M$ given by (\ref{three}), becomes
\begin{equation}
M \, = \, (R_{AdS}^2/r^3)M^2 L_P^2 \, ,
\end{equation}
which nicely leads to the mass-to-size relation 
\begin{equation}
M \, = \, r^3/(R_{AdS}^2L_P^2) \, . 
\label{massrelation}
\end{equation}
This is a well-known mass-to-size relation for a large black hole in AdS!

Now we can understand the deep meaning of the number defined in (\ref{two}). It is simply the 
Bekenstein-Hawking entropy of the large black hole.  Namely,  after plugging the above expression for the mass in (\ref{two}) we get
\begin{equation}
N(r) \,  \equiv \,  (R_{AdS}/r)^4N_M \, = \, r^2/L_P^2 \,, 
\end{equation}
which is nothing else but the Bekenstein Hawking entropy.

The graviton portrait of large black holes in AdS also sheds some interesting light on the corresponding quantum wave function. Indeed the natural quantum state for the combined system of gravitons should be a pure state in the tensor product of the Hilbert spaces of the two types of gravitons
\begin{equation}
|\Psi \rangle \,  =  \, \sum |\Psi_{AdS}\rangle  \otimes |\Psi_{M} \rangle \, .
\end{equation}
At criticality we expect this state to be maximally entangled. As discussed above criticality takes place when the gravitons sourced by $M$ are blue shifted and its number is reduced to $N$ as defined in equation (\ref{two}). In the spirit of thermofield formalism \cite{Jap} \cite{Malda2} \cite{Pap} we can use this maximally entangled pure state to define a reduced thermal density matrix. If state is maximally-entangled the Von Neumann entropy will be given by $S \, \sim \,$ log$(d)$ where $d$ is the dimension of the smaller Hilbert space. At criticality the number of degrees of freedom are respectively $N$ and $N_{AdS}$. Since for large black holes $N\, < \, N_{AdS}$, the entropy for maximal entanglement is $S\sim N$. Therefore the reduced density matrix approximately describes a thermal state at temperature $T \, \sim \, M\, S$,  i.e., at the large black hole Hawking temperature.

The case of small black holes $r \, \ll \, R_{AdS}$ is physically simpler.  
 In this case the gravitons sourced by $M$ are harder than those sourced by the cosmological constant and the 
 effect of the latter in the regime $r \, \ll \, R_{AdS}$ is negligible. In other words, the small black holes in AdS do  not differ significantly from  asymptotically flat black holes. In summary,  we have been able to describe a great amount of the known physics of black holes in AdS in terms of graviton condensates at criticality. 

\section{Conclusions and Outlook} 

 In this paper we have pushed further the compositeness approach to gravity put forward 
  in \cite{Nportrait, Quantum}. 
  The key concept is to replace the {\it classical}  curved background metric 
   by the {\it quantum} notion of a composite multi-particle state built on Minkowski vacuum. The role of the quantum constituents is played by the longitudinal (usually off-shell) gravitons, with the characteristic wave-length and/or frequencies set by a would-be classical curvature radius. 
   
    For concrete computations, we have adopted the 
particle number representation $|N\rangle \, \equiv \, (a^{+})^N |0\rangle$ of states that in $N = \infty$ recover the classical geometry. In this way,  a curved background is represented by a quantum state in a Fock space of off-shell gravitons.  Alternatively, we can use the coherent state representation, which 
in $N=\infty$ limit gives similar results. 
  
   For maximally-symmetric spaces, such as black holes,  de Sitter or AdS , the physics of such 
 representation is very similar to a Bose-Einstein condensate at the quantum critical point.   
   
     The known phenomena available in the metric 
    description, such as the geodesic motion and/or the particle-creation in a background metric, 
can be recovered as the $N=\infty$ limit of quantum scattering processes of the constituent gravitons.  
 The geometric interpretation emerges in the infinite-$N$ limit from  transition  matrix elements of the type: $\langle N-1| a |N \rangle\,,~~ \langle N| a^+a |N \rangle\\ ,..$ .  In particular we have illustrated this emergence for the examples of  de Sitter and AdS spaces. 
 
   One important aspect of the composite picture is that particle-creation is not a vacuum process, 
 but rather the result of scattering of the initially-existing off-shell gravitons during which some of them 
 become on-shell final states. 
  This process, can be ``confused" with the vacuum process only  for  $N = \infty$. This fact 
  fundamentally changes the physics of entanglement-generation in the black hole radiation giving a new twist to the information paradox discussion.    
  
   Applying this framework  to inflationary spaces, we have recovered the known predictions in simple terms
of scattering of the constituent gravitons and inflatons. However, since the particle-creation is not a vacuum process, 
but rather originates from the composite structure of the background,  
it imprints some cumulative  effects in the cosmological observables that scan the entire history of the inflationary patch. 
This creates a (so far theoretical) hope of detecting the imprints of  the Universe's history dating way before the 
last $60$ e-foldings prior to the end of inflation.    
    
   Next, compositeness imposes a severe bound on the duration of the inflationary and/or de Sitter states, beyond which 
   none of these spaces can be treated as approximately-classical. The bound is non-perturbative and cannot 
   be removed by some re-summation.  This raises the question about the eternity of such states.  What we are discovering is that de Sitter and inflationary spaces cannot be eternal in the usually-used sense of eternity-valid in the approximate classical description of the background.  However, as we have discussed, they could in principle persist eternally in a new  quantum state, no longer subjected to a 
   classical metric description.  
   
     The quantum physics of de Sitter that we are uncovering could provide an answer to 
   the issues raised in \cite{Tsam} and \cite{Polya}.  Some time ago these authors have pointed out certain 
   instabilities appearing within the semi-classical treatment of de Sitter background, which could be interpreted as the problems with the eternal de Sitter.  However, within semi-classical treatment it is impossible to fully 
 clarify the meaning of these instabilities, since they could, for example, be attributed to a breakdown of the perturbative expansion.  Our picture provides a concrete framework in which the issues raised by these authors can (in principle) be answered. In particular,  we see that the 
 effects cannot be blamed on breakdown of perturbation theory and are real.  Most importantly, we can identify 
 their microscopic physical meaning. 
   
  Indeed, in our description de Sitter, within a finite time, becomes a fully quantum 
 entity, in which order half of the constituent gravitons have been depleted and red-shifted away. 
 It is not excluded that the system can get stabilized (or at least become very long lived) in such a 
 maximally-entangled and half-depleted quantum state.  As we have discussed above, presumably what replaces in such a state
 the classical characteristics, is the quantum relation $\dot{N} \, = \, 0$. 
  Understanding the underlying nature and consistency of 
 such a state is crucial for addressing the cosmological constant problem.  Whatever is the outcome of this study,  it is evident that the cosmological constant problem will acquire a new appearance. 
 
  For instance,  
proving inconsistency of such a quantum state, would of course provide a strong evidence of 
incompatibility of the positive cosmological constant with quantum physics. 
On the other hand, if such a quantum state is proved to be consistent, the situation will become  
very subtle and we will need to reconsider the way we pose the problem. 
In any case, this is an interesting problem for future investigation. 

 Finally, our picture provides a quantum foundation of what can be called the emergence of holography.  
 In particular, it establishes an explicit connection 
 between the (seemingly) classical gravitational backgrounds  and approximately conformal quantum 
theories.  In our description the two appear as parts of the same entity:  The Bose-gas of gravitons 
at the quantum critical point.  The mean-field description of this Bose-Einstein condensate represents 
the classical metric, whereas  the  approximate CFT is a theory governing collective Bogoliubov 
 modes of the condensate that become nearly gapless at the critical point.
 With this logic, treating AdS space as a graviton Bose-gas at the critical point, the CFT emerges as an effective 
 description of the Bogoliubov modes of the graviton condensate.  The coincidence between the 
 parameters of this system and the central charge of CFT is remarkable. 
   At this point we cannot claim that 
 this is the origin of AdS/CFT conjecture, but similarities are striking and deserve further investigation. 
  
   Our observations set possible lines of investigation in several different  directions.
 In particular, as suggested in \cite{class1},\cite{Nportrait},\cite{P7}, it is natural to generalize compositeness approach to non-gravitational classical entities, such as solitons  and other non-perturbative field configurations.  
 For example, as shown in \cite{Nportrait} the counting of constituent soft gauge bosons in the field of magnetic  
 monopoles is very similar to counting of longitudinal gravitons in a black hole, with the role of the Planck mass
 taken up by the scalar vacuum expectation value.  Other field configurations can be treated in exactly the same spirit. 
  The compositeness effects than will show up as new $1/N$-effects, are not accounted by the standard perturbative analysis.

\section*{Acknowledgements}

 It is a pleasure to thank Jose Barbon,  Daniel Flassig, Valentino Foit,  Andre Franca,  Tehseen Rug,
 Alex Pritzel and Nico Wintergerst for discussions.   
The work of G.D. was supported in part by Humboldt Foundation under Alexander von Humboldt Professorship,  by European Commission  under 
the ERC Advanced Grant 339169,   by TRR 33 \textquotedblleft The Dark
Universe\textquotedblright\   and  by the NSF grant PHY-0758032. 
The work of C.G. was supported in part by Humboldt Foundation, by Grants: FPA 2009-07908, CPAN (CSD2007-00042) and HEPHACOS P-ESP00346 and by the ERC Advanced Grant 339169.

\end{document}